\documentclass[prd,aps,showpacs,notitlepage,superscriptaddress,nofootinbib]{revtex4-1}
\usepackage{amsmath}
\usepackage{epsfig}
\def\bea#1\eea{\begin{align}#1\end{align}} 
\newcommand{\nnu}{\nonumber\\}
\newcommand{\bef}{\begin{figure}[htb]\centering}
\newcommand{\eef}{\end{figure}}
\newcommand{\hx}{\hat{x}}
\newcommand{\hz}{\hat{z}}
\allowdisplaybreaks

\begin{document}
\title{Transverse momentum broadening in semi-inclusive deep inelastic scattering \\ at next-to-leading order}

\date{\today}

\author{Zhong-Bo Kang}
\email{zkang@lanl.gov}
\affiliation{Theoretical Division, 
                   Los Alamos National Laboratory, 
                   Los Alamos, NM 87545, USA}

\author{Enke Wang}
\email{wangek@mail.ccnu.edu.cn}
\affiliation{Institute of Particle Physics and Key Laboratory of Lepton and Quark Physics (MOE), 
                   Central China Normal University, 
                   Wuhan 430079, China}

\author{Xin-Nian Wang}
\email{xnwang@lbl.gov}
\affiliation{Institute of Particle Physics and Key Laboratory of Lepton and Quark Physics (MOE),
                  Central China Normal University, 
                  Wuhan 430079, China}
\affiliation{Nuclear Science Division, 
                   Lawrence Berkeley National Laboratory, 
                   Berkeley, CA 94720, USA}

\author{Hongxi Xing}
\email{hxing@northwestern.edu}
\affiliation{High Energy Physics Division, Argonne National Laboratory, Argonne, IL 60439, USA}
\affiliation{Department of Physics and Astronomy, Northwestern University, Evanston, IL 60208, USA}
\affiliation{Theoretical Division, Los Alamos National Laboratory, Los Alamos, NM 87545, USA}

\begin{abstract}
Within the framework of higher-twist collinear factorization, transverse momentum broadening for the final hadrons in semi-inclusive deeply inelastic $e+A$ collisions is studied at the next-to-leading order (NLO) in perturbative QCD. Through explicit calculations of real and virtual corrections at twist-4, the transverse-momentum-weighted differential cross section due to double scattering is shown to factorize at NLO and can be expressed as a convolution of twist-4 nuclear parton correlation functions, the usual twist-2 fragmentation functions and hard parts which are finite and free of any divergences. A QCD evolution equation is also derived for the renormalized twist-4 quark-gluon correlation function which can be applied to future phenomenological studies of transverse momentum broadening and jet quenching at NLO.
\end{abstract}


\maketitle

\section{Introduction}
Multiple scatterings in deeply inelastic lepton-nucleus scattering, hadron-nucleus and heavy-ion collisions lead to many
interesting phenomena that in turn can provide useful tools for diagnosing properties of cold and hot nuclear media \cite{Gyulassy:2003mc, Abreu:2007kv, Salgado:2011wc, Albacete:2013ei}. The predicted phenomena such as  jet quenching and transverse momentum broadening \cite{Gyulassy:1993hr, Baier:1996sk,Gyulassy:2000er,Zakharov:1997uu,Wang:2001ifa,Guo:2000nz,Vitev:2002pf, Kang:2012kc} have been observed in the fixed target experiments at the Deutsches Elektronen-Synchrotron, Jefferson Lab and Fermilab \cite{Airapetian:2007vu,Airapetian:2009jy,Brooks:2011sa,Accardi:2009qv,McGaughey:1999mq,Peng:1999gx,Johnson:2006wi,Leitch:2005yi}, as well as in the ongoing collider experiments at Relativistic Heavy Ion Collider and the Large Hadron Collider \cite{Muller:2012zq,Majumder:2010qh}. These phenomena will continue to be the focus of future studies in experiments at the proposed Electron Ion Collider \cite{Accardi:2012qut,Boer:2011fh}. 

Many theoretical formalisms have been developed in the study of multiple scatterings which in turn can be used to extract medium properties from the nontrivial nuclear dependence observed in the experiments of high-energy collisions with nuclear targets. Significant progress has been made in the past few years, in particular, in the study of parton energy loss \cite{Liou:2014rha,Wu:2014nca,Huang:2013vaa,Kang:2014xsa,Chien:2015vja}, radiative corrections to transverse momentum broadening \cite{Mueller:2012bn,Liou:2013qya,Iancu:2014kga,Blaizot:2014bha}, effects of multiple gluon emissions \cite{MehtarTani:2010ma,Blaizot:2013hx,Fickinger:2013xwa}, and phenomenological extraction of the jet transport parameter from jet quenching in high-energy heavy-ion collisions \cite{Burke:2013yra}.  However, so far a complete next-to-leading-order (NLO) calculation in perturbative QCD (pQCD) for either jet quenching or transverse momentum broadening is still lacking, which is essential for more precise extraction of medium properties from experimental data. 

One of the approaches in studying effects of multiple scatterings is based on a generalized high-twist factorization theorem \cite{Luo:1992fz,Luo:1994np, Qiu:1990xy,Collins:1989gx}. Within such an approach, these effects manifest themselves as power corrections to the differential cross sections, whose main contributions often depend on high-twist matrix elements of the nuclear state that are enhanced by the nuclear size. So far most studies have focused on double parton scatterings and their effect on transverse momentum broadening, which leads to nuclear enhancement in the dijet transverse momentum imbalance in photon-nucleus collisions \cite{Luo:1993ui}, transverse momentum broadening for single inclusive jet (or hadron) production in semi-inclusive deep inelastic scattering (SIDIS) \cite{Guo:2000eu,Majumder:2007hx,Guo:1998rd}, as well as Drell-Yan lepton pair~\cite{Guo:1998rd,Fries:2002mu}, vector-boson \cite{Kang:2008us,Kang:2012am} and back-to-back particle productions \cite{Kang:2011bp,Xing:2012ii} in $p+A$ collisions. Phenomenological studies of experimental data within the high-twist formalism have been quite successful \cite{Luo:1993ui,Guo:1998rd,Kang:2008us,Kang:2011bp,Adam:2015jsa}, giving us confidence in using multiple scatterings and the phenomenology as a tool to probe the fundamental twist-4 nuclear parton correlation functions and the associated QCD dynamics. All these calculations, however, are based on the picture of the leading-order (LO)  ``bare'' twist-4 factorization without higher-order corrections in pQCD and contribute to most of the theoretical uncertainties in the phenomenological studies of jet quenching \cite{Armesto:2011ht}. More complete NLO calculations with renormalized twist-4 matrix elements and finite corrections are very complex and have not been attempted so far.  They are, however, necessary for more accurate predictions and more precise extraction of medium properties from future phenomenological studies of experimental data.

In this paper, we carry out the calculation of the one-loop perturbative corrections to the transverse-momentum-weighted SIDIS cross section at twist 4. Through explicit calculations, we will illustrate the factorization of the transverse-momentum-weighted cross section of SIDIS and derive the evolution equations for the renormalized twist-4 two-parton correlation functions. Such a calculation is an important step towards a full NLO  pQCD description of single hadron spectra in SIDIS and transverse momentum broadening within the high-twist formalism. A brief summary of our results has been reported earlier in Refs. \cite{Kang:2013raa,Xing:2014kpa}. We now provide detailed derivations and discussions in this paper. The rest of our paper is organized as follows. In Sec. II, we introduce our notations and kinematics, and review the LO derivation for transverse momentum broadening. In Sec. III, we perform explicit calculations of NLO contributions at twist 4 to the transverse-momentum-weighted differential cross section, including quark-gluon and gluon-gluon double scatterings, as well as the interference contributions from single and triple scatterings. In particular, we show the complete cancellation of soft divergences in real and virtual corrections.  The remaining collinear divergences can be absorbed into the standard fragmentation function and/or the twist-4 parton correlation function of the nuclear state, which give rise to the factorization scale evolution of these functions. We summarize our paper in Sec. IV.

\section{Transverse momentum broadening at leading order }

\subsection{Notations and kinematics}
We start this section by specifying our notations and kinematics in SIDIS. We consider a lepton $l$ scattering off a large nucleus $A$, 	
\bea
l(L_1)+A(p) \rightarrow l(L_2)+h(\ell_h)+X,
\eea
where $L_1$ and $L_2$ are the four-momenta of the incoming and outgoing leptons, $\ell_h$ is the observed hadron momentum, and $p$ is the momentum per nucleon in the nucleus with the atomic number $A$. In the approximation of one-photon exchange, the virtual photon momentum is given by $q=L_1-L_2$ with the invariant mass $Q^2=-q^2$. The usual SIDIS Lorentz-invariant variables are defined as follows:
\bea
x_B=\frac{Q^2}{2p\cdot q},\qquad
y=\frac{p\cdot q}{p\cdot L_1}, \qquad
z_h=\frac{p\cdot\ell_h}{p\cdot q}.
\eea
For later convenience, we also define Mandelstam variables at the partonic level,
\bea
\hat s=(xp+q)^2,\qquad
\hat t=(\ell-q)^2,\qquad
\hat u=(\ell-xp)^2,
\eea
where $\ell$ is the momentum of the final-state parton which fragments into the observed hadron $h$. It is instructive to realize that the transverse momentum $\ell_T$ of the final-state parton in the so-called {\it hadron frame} \cite{Meng:1991da,Kang:2008qh} can be written in terms of Mandelstam variables as
\bea
\ell_{T}^2=\frac{\hat s\hat t\hat u}{(\hat s+Q^2)^2}.
\eea

The transverse momentum broadening, 
\bea
\Delta \langle\ell_{hT}^2\rangle = \langle\ell_{hT}^2\rangle_{eA}-\langle\ell_{hT}^2\rangle_{ep}, 
\label{definition}
\eea
is defined as the difference between the average squared transverse momentum of the observed hadron produced on a nuclear target ($e+A$ collisions) and that on a proton target ($e+p$ scattering), with $\langle\ell_{hT}^2\rangle$ given by
\bea
\langle\ell_{hT}^2\rangle = \int d\ell_{hT}^2 \ell_{hT}^2 \frac{d\sigma}{d{\cal PS}d\ell_{hT}^2}\left/\frac{d\sigma}{d{\cal PS} }\right. ,
\label{pht-weight}
\eea
where the phase space $d{\cal PS} = dx_Bdydz_h$. The denominator gives the so-called single hadron differential cross section in SIDIS, which can be written as \cite{Graudenz:1994dq,Daleo:2004pn}
\bea
\frac{d\sigma}{d{\cal PS}}=\frac{\alpha_{em}^2}{Q^2}\left[Y^M(-g^{\mu\nu})+Y^L\frac{4x_B^2}{Q^2}p^{\mu}p^{\nu}\right]\frac{dW_{\mu\nu}}{dz_h},
\label{eq-cross}
\eea
where $\alpha_{em}$ stands for the fine-structure constant, and $W_{\mu\nu}$ is the hadronic tensor for $\gamma^*+A \to h +X$. Here the term proportional to $Y^M$ is the so-called ``metric'' contribution, while the term proportional to $Y^L$ is the longitudinal contribution.   $Y^M$  and $Y^L$ are closely connected to the photon polarizations with the following expressions,
\bea
Y^M=\frac{1+(1-y)^2}{2y}, \qquad 
Y^L=\frac{1+4(1-y)+(1-y)^2}{2y}.
\eea

The result for single hadron differential cross section in SIDIS at leading twist is well-known. As a warm-up exercise, we also calculate this cross section to NLO. Working in $n=4-2\epsilon$ dimensions with the $\overline{\rm MS}$ scheme, our findings are consistent with those in the literature~\cite{Altarelli:1979ub, Kang:2012ns,VITEV:2014wza,Anderle:2012rq}:
\bea
\frac{d\sigma}{d{\cal PS}}
=&\sigma_0\sum_q e_q^2
\int \frac{dx}{x} \frac{dz}{z} f_{q/A}(x, \mu_f^2) D_{h/q}(z, \mu_f^2) \delta(1-\hat x)\delta(1-\hat z)
\nnu
&
+\sigma_0 \frac{\alpha_s}{2\pi} \sum_q e_q^2 \int \frac{dx}{x} \frac{dz}{z} 
 f_{q/A}(x, \mu_f^2) D_{h/q}(z, \mu_f^2)
\Bigg\{
\ln\left(\frac{Q^2}{\mu_f^2}\right)
\left[P_{qq}(\hx)\delta(1-\hz)+ P_{qq}(\hz)\delta(1-\hx) \right]
+H^{NLO}_{T2-qq}\Bigg\}
\nnu
&
+\sigma_0 \frac{\alpha_s}{2\pi} \sum_q e_q^2 \int \frac{dx}{x} \frac{dz}{z} 
 f_{q/A}(x, \mu_f^2) D_{h/g}(z, \mu_f^2)
\Bigg[
\ln\left(\frac{Q^2}{\mu_f^2}\right)
P_{gq}(\hz)\delta(1-\hx)
+H^{NLO}_{T2-qg}
\Bigg]
\nnu
&+\sigma_0 \frac{\alpha_s}{2\pi} \sum_q e_q^2 \int \frac{dx}{x} \frac{dz}{z} 
f_{g/A}(x, \mu_f^2) \left[ D_{h/q}(z, \mu_f^2)+D_{h/\bar q}(z, \mu_f^2)\right]
\left[ \ln\left(\frac{Q^2}{\mu_f^2}\right)P_{qg}(\hx)\delta(1-\hz)+H^{NLO}_{T2-gq}\right], 
\label{eq-single}
\eea
where $\mu_f$ is the factorization scale, $f_{q(g)/A}(x,\mu_f^2)$ is the quark (gluon) distribution function inside the nucleus, and $D_{h/q(g)}(z,\mu_f^2)$ is the fragmentation function for a quark (gluon) into a hadron $h$. The detailed expressions for the finite terms $H^{NLO}_{T2-qq}$, $H^{NLO}_{T2-qg}$, and $H^{NLO}_{T2-gq}$ are given in Appendix by Eqs.~\eqref{eq-finiteT2-qq}, \eqref{eq-finiteT2-qg}, and \eqref{eq-finiteT2-gq}, respectively. Other variables are defined as $\hat x=x_B/x$, $\hat z=z_h/z$, and $\sigma_{0}$ is given by
\bea
\sigma_{0}=\frac{2\pi\alpha_{em}^2}{Q^2}\frac{1+(1-y)^2}{y}(1-\epsilon).
\label{eq-sigma0}
\eea
In Eq.~\eqref{eq-single}, $P_{ab}(z)$ is the usual Dokshitzer-Gribov-Lipatov-Altarelli-Parisi (DGLAP) splitting kernel for partons $b\to a$
\bea
P_{qq}(z) &= C_F \left[\frac{1+z^2}{(1-z)_+}+\frac{3}{2} \delta(1-z) \right],
\label{Pqq}
\\
P_{gq}(z) &= C_F \frac{1+(1-z)^2}{z},
\\
P_{qg}(z) &= T_R \left[z^2+(1-z)^2\right],
\label{Pqg}
\eea
where $C_F=(N_c^2-1)/2N_c$ with $N_c=3$ being the number of colors, and $T_R=1/2$.

\subsection{Transverse momentum broadening:  Leading order}
\bef
\psfig{file=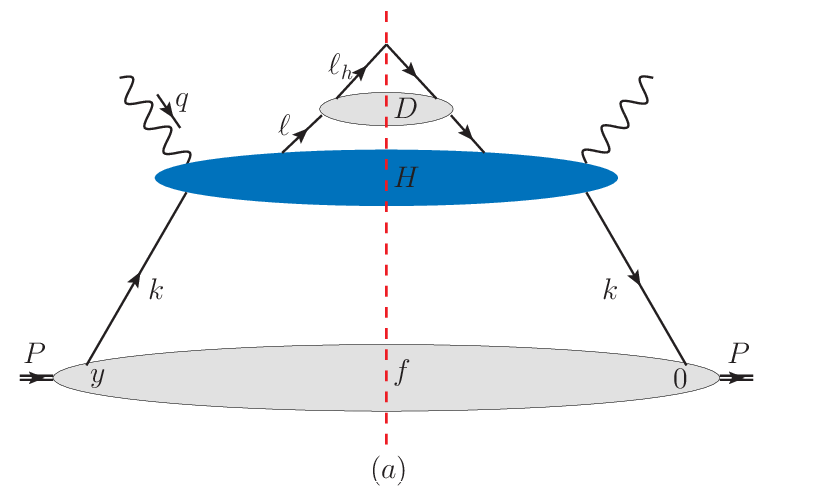, width=2.5in}
\hskip 0.2in
\psfig{file=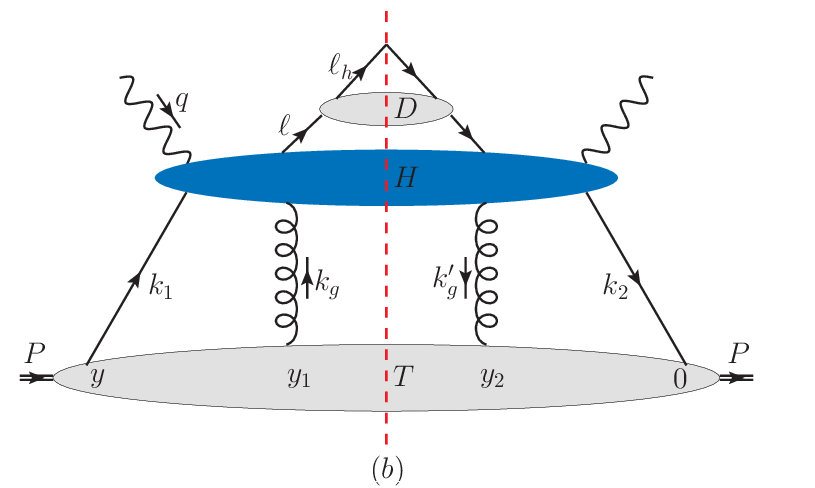, width=2.5in}
\caption{The general diagrams for single inclusive hadron production in SIDIS in a nuclear medium: (a) single scattering contribution with $k=x\,p$; (b) quark-gluon double scattering with the initial parton's momenta $k_1=x_1p$, $k_2=(x_1+x_3)p+k_{2T}-k_{3T}$, $k_g=x_2p+k_{2T}$ and $k_g'=(x_2-x_3)p+k_{3T}$, respectively. Here $k_T$ is the transverse momentum kick from the nucleus.}
\label{fig-DIS}
\eef
In a nuclear medium, the outgoing parton in SIDIS may experience additional scatterings with other partons from the nucleus before fragmenting into final observed hadrons. Taking into account these multiple scatterings, one can express the  differential cross section for single inclusive hadron production in SIDIS off a nuclear target as a sum of contributions from single, double, and higher multiple scatterings \cite{Qiu:2001hj},
\bea
d\sigma = d\sigma^{S}+d\sigma^{D}+\dots \, ,
\label{single-double}
\eea
where the superscript ``$S$'' (``$D$'') indicates the single (double) scattering contribution. In the case of a single scattering as illustrated in Fig.~\ref{fig-DIS}(a), the virtual photon interacts with a single parton (quark or gluon) coming from the nucleus to produce a parton which will then fragment into the final observed hadron. Such a single scattering is localized in space and time, and thus usually does not lead to a significant modification for the production rate from $e+p$ to $e+A$ collisions, except for a mild $A$ dependence from nuclear parton distribution functions. On the other hand, in the case of double scatterings as shown in Fig.~\ref{fig-DIS}(b), the outgoing parton experiences one additional scattering with another parton (e.g., a gluon in the figure) from the nucleus. Such a double scattering is usually power suppressed $\sim 1/Q^2$ by the hard scale of the process, though it could be enhanced by the nuclear size $\sim A^{1/3}$, which happens when two partons come from different nucleons inside the nucleus. In this situation, double scatterings will then lead to a nuclear enhancement in the average squared transverse momentum for the observed hadron produced in $e+A$ collisions. Even higher multiple scattering will be suppressed even more by the hard scale $Q$~\footnote{This statement might not be necessarily true for the small Bjorken-$x$ limit, where gluon density in the target is extremely high and thus all the multiple scatterings are equally important and have to be resummed. In such a kinematic region, we have to rely on a different theoretical framework, see e.g. \cite{Gelis:2010nm}.}, and thus the leading contribution to the transverse momentum broadening $\Delta\langle \ell_{hT}^2\rangle$ should come from the double scattering. According to Eqs.~\eqref{definition} and \eqref{single-double}, the leading contribution to the defined nuclear transverse momentum broadening can thus be written as,
\bea
\Delta \langle\ell_{hT}^2\rangle \approx \frac{d\langle \ell_{hT}^2 \sigma^D\rangle}{d{\cal PS}} \left/\frac{d\sigma}{d{\cal PS}}\right. ,
\qquad
\frac{d\langle \ell_{hT}^2 \sigma^D\rangle}{d{\cal PS}} 
\equiv \int d\ell_{hT}^2 \ell_{hT}^2 \frac{d\sigma^D}{d{\cal PS}d\ell_{hT}^2},
\label{eq-broadening definition}
\eea
where the superscript ``$D$'' indicates the double-scattering contribution. 

It is instructive to emphasize that single scattering certainly contributes in the calculation of $\langle \ell_{hT}^2\rangle$ for both $e+A$ and $e+p$ collisions. Such a contribution $\sim  \int d\ell_{hT}^2 \ell_{hT}^2 d\sigma^S$ can even produce a divergent result, because of the high-$\ell_{hT}$ perturbative tail of the cross section $d\sigma^S$~\cite{Boer:2015kxa}. However, such contributions do not affect our analysis, since we are studying the transverse momentum broadening $\Delta \langle\ell_{hT}^2\rangle$ of Eq.~\eqref{definition}, which is defined as the difference in $\langle \ell_{hT}^2\rangle$ between $e+A$ and $e+p$ collisions. Those divergences thus cancel in the final result and we end up with physically meaningful results. In other words, the transverse momentum broadening is closely related to the transverse momentum $\ell_{hT}^2$-weighted  differential cross section through double scattering, i.e., the numerator in the left equation of Eq.~\eqref{eq-broadening definition}, which will be the focus of our paper. It is worth mentioning that the contributions from double scatterings are related to the so-called power corrections at twist-4 level, for which a justification of a generalized factorization formalism was given in Refs.~\cite{Qiu:1990xy,Kang:2014tta}. Our computations in the current paper can be regarded as a verification of such a factorization formalism up to one-loop order. 
\bef
\psfig{file=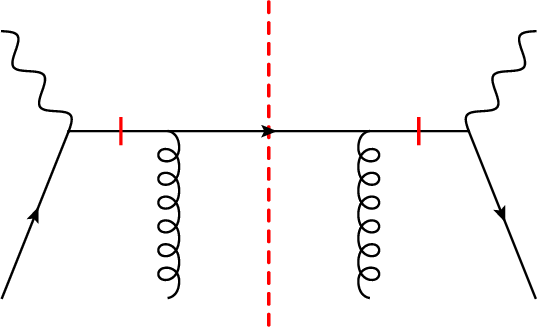, width=1.8in}
\caption{Feynman diagram for the double-scattering contribution to transverse momentum broadening at leading order. The parton momenta follow the same notation as in Fig.~\ref{fig-DIS} (b). The short bars indicate the propagators where the soft poles arise.}
\label{fig-LO}
\eef

At leading order, the double-scattering contribution is given by Fig.~\ref{fig-LO}. The $\ell_{hT}^2$-weighted cross section (specifically the hadronic tensor) can be written as
\bea
\frac{d\langle\ell_{hT}^2W^D\rangle}{dz_h}^{\rm (LO)}=&\frac{2\pi x_B}{Q^2}\sum_qe_q^2\int\frac{dx}{x}\frac{dz}{z}D_{h/q}(z)
\int d^{n-2}\ell_{hT}
\,\ell_{hT}^2\, \delta^{n-2}(\ell_{hT}-zk_T)\delta(1-{\hat x})\delta(1-{\hat z})
\nnu
&\times\int d^{n-2}k_T dx_1dx_2dx_3T_A(x_1,x_2,x_3,k_T)(-g^{\mu\nu})H_{\mu\nu}\left(\{x_i\},p,q,\ell,\ell_h,k_T\right) \delta(x_1+x_2-x),
\label{eq-LO1}
\eea
where $W^D=(-g^{\mu\nu})W^D_{\mu\nu}$. In arriving at this, we have used the fact that the longitudinal contribution vanishes at LO, and thus the only contribution is the metric term. In Eq.~\eqref{eq-LO1}, $\{x_i\} = \{x_1,x_2,x_3\}$ are the independent collinear momentum fractions carried by partons from the nucleus, $k_T$ is a small transverse momentum kick due to the multiple scattering, and the matrix element $T_A(x_1,x_2,x_3,k_T)$ is defined as
\bea
T_A(x_1,x_2,x_3,k_T)=&\int\frac{dy^-}{2\pi}\frac{dy_1^-}{2\pi}\frac{dy_2^-}{2\pi}\frac{d^2y_T}{(2\pi)^2}
e^{ix_1p^+y^-}e^{ix_2p^+(y_1^--y_2^-)}e^{ix_3p^+y_2^-}e^{ik_T\cdot y_{T}}
\nnu
&\times
\frac{1}{2}\langle A|\bar{\psi}_q(0)\gamma^+A^+(y_2^-,0_T)A^{+}(y_1^-,y_T)\psi_q(y^-)|A\rangle.
\label{eq-TA}
\eea
As in Refs.~\cite{Kang:2008us,Kang:2011bp,Xing:2012ii}, the calculation proceeds by first taking the Taylor expansion of the hard part 
function in $k_T$,
\bea
H_{\mu\nu}\left(\{x_i\},p,q,\ell,\ell_h,k_T\right) = H_{\mu\nu}\left(\{x_i\},p,q,\ell,\ell_h,k_T=0\right) + {\cal O}_{\mu\nu}(k_T^2).
\eea
Note that the term linear in $k_T$ in the above expansion does not contribute to unpolarized SIDIS.  Using $\delta^{n-2}(\ell_{hT}-zk_T)$ in Eq.~\eqref{eq-LO1} to set $\ell_{hT}=z k_T$, one can convert $k_T^2A^+A^+$ to the gauge-covariant gluon field strength $F^+_{\sigma}F^{\sigma+}$ in the matrix element through partial integrations in $y_T$.  We further integrate over the momentum fractions $x_1, x_2, x_3$ through contour 
integrations around poles in the hard part $H_{\mu\nu}$,
\bea
\frac{1}{(q+x_1p)^2+i\epsilon}&=\frac{x_B}{Q^2}\frac{1}{x_1-x_B+i\epsilon}, 
\\
\frac{1}{\left[q+(x_1+x_3)p\right]^2-i\epsilon}&=\frac{x_B}{Q^2}\frac{1}{x_1+x_3-x_B-i\epsilon}.
\eea
Together with the phase space $\delta$ function $\delta(x_1+x_2 - x_B)$ in Eq.~\eqref{eq-LO1}, we fix $x_1=x_B$, $x_2=0$, and $x_3=0$. Finally we have
\bea
\frac{d\langle\ell_{hT}^2W^D\rangle}{dz_h}^{\rm (LO)}
= \frac{2\alpha_s}{N_c}z_h^2(2\pi)^3(1-\epsilon)\sum_qe_q^2\int\frac{dx}{x}
T_{qg}(x,0,0)\int\frac{dz}{z}D_{h/q}(z)\delta(1-{\hat x})\delta(1-{\hat z}),
\eea
where the twist-4 quark-gluon correlation function $T_{qg}(x_1, x_2, x_3)$ is given by \cite{Wang:2001ifa,Guo:2000nz,Kang:2008us} \footnote {Our notation here follows Refs. \cite{Luo:1992fz,Luo:1994np}, which differs by $1/2\pi$ as compared to Refs. \cite{Wang:2001ifa,Guo:2000nz}.},
\bea
T_{qg}(x_1, x_2, x_3)
=&\int \frac{dy^-}{2\pi} e^{ix_1p^+y^-}  \int \frac{dy_1^-dy_2^-}{4\pi} e^{ix_2p^+(y_1^- - y_2^-)}
 e^{ix_3p^+y_2^-} \theta(y_2^-)\theta(y_1^- - y^-)
\nnu
&
\times  \langle A|{\bar\psi}_q(0) \gamma^+ F_{\sigma}^+(y_2^-)F^{\sigma +}(y_1^-)\psi_q(y^-)|A\rangle.
\label{Tqg}
\eea
We thus obtain the double-scattering contribution to the $\ell_{hT}^2$-weighed differential cross section at LO,
\bea
\frac{d\langle\ell_{hT}^2\sigma^D \rangle^{\rm (LO)}}{d{\cal PS}}=&\sigma_h\sum_qe_q^2\int\frac{dx}{x}T_{qg}(x,0,0)\int\frac{dz}{z}D_{h/q}(z)\delta(1-{\hat x})\delta(1-{\hat z}),
\label{LO-weight}
\eea
where $\sigma_h=(4\pi^2\alpha_sz_h^2/N_c)\sigma_0$, with $\sigma_0$ defined in Eq. (\ref{eq-sigma0}). The LO transverse momentum broadening is then 
\bea
\Delta\langle\ell_{hT}^2\rangle=\left(\frac{4\pi^2\alpha_sz_h^2}{N_c}\right)\frac{\sum_qe_q^2T_{qg}(x_B,0,0)D_{h/q}(z_h)}{\sum_qe_q^2f_{q/A}(x_B)D_{h/q}(z_h)},
\eea
as obtained in previous calculations \cite{Guo:2000eu,Guo:1998rd}.

\section{Transverse momentum broadening at next-to-leading order}
In this section, we present our calculations of NLO contributions to transverse momentum broadening in SIDIS. We first study the virtual-photon-quark ($\gamma^*+q$) interaction channel, which involves the quark-gluon correlation function $T_{qg}$ as defined in Eq.~\eqref{Tqg}. We then derive the result for the virtual-photon-gluon ($\gamma^*+g$) channel, which involves the gluon-gluon correlation function $T_{gg}$ defined in Eq.~\eqref{eq-Tgg} below. The final result will be presented at the end of this section. 

The double-scattering contributions in the nuclear medium manifest themselves as power corrections to the differential cross section.  A high-twist factorization formalism was established \cite{Luo:1992fz,Luo:1994np,Qiu:1990xy} to systematically extract these contributions. This formalism stems directly from the well-established collinear factorization theorem \cite{Luo:1992fz,Luo:1994np, Qiu:1990xy, Collins:1989gx} and has recently been extended to include transverse-momentum-dependent parton distributions \cite{Liang:2006wp,Song:2010pf,Gao:2010mj}. Within such an approach, one carries out a collinear expansion of hard parts and reorganizes the final results in terms of power corrections, where the second-order expansion gives rise to the twist-4 contribution. In the presence of a large nucleus ($A\gg 1$), the dominant contribution comes from the terms associated with the high-twist matrix elements of the nuclear state that are enhanced by the nuclear size. The general formalism for the double-scattering contribution can be written as
\bea
\frac{dW_{\mu\nu}^D}{dz_h}=&\sum_qe_q^2\int\frac{dz}{z}D_{h/q}(z)\int\frac{dy^-}{2\pi}\frac{dy_1^-}{2\pi}\frac{dy_2^-}{2\pi}
\frac{1}{2}\langle A|\bar{\psi}_q(0)\gamma^+F_{\sigma}^+(y_2^-)F^{\sigma +}(y_1^-)\psi_q(y^-)|A\rangle \nnu
&\times \left[-\frac{1}{2(1-\epsilon)}g^{\alpha\beta}\right]\left[\frac{\partial^2}{\partial k_{2T}^{\alpha}\partial k_{3T}^{\beta}}
{\overline H}_{\mu\nu}(p,q,\ell,\ell_h,k_{2T},k_{3T},\{y_i\})\right]_{k_{2T}=k_{3T}=0},
\label{eq-HTgeneral}
\eea
where $\{y_i\}=\{y,y_1,y_2\}$, and ${\overline H}_{\mu\nu}(p,q,\ell,\ell_h,k_{2T},k_{3T},\{y_i\})$ is the Fourier transform of the hard partonic function $H_{\mu\nu}\left(\{x_i\},p,q,\ell,\ell_h,k_{2T},k_{3T},\{y_i\}\right)$,
\bea
{\overline H}_{\mu\nu}(p,q,\ell,\ell_h,k_{2T},k_{3T},\{y_i\})=\int dx_1dx_2dx_3e^{ix_1p^+y^-}e^{ix_2p^+(y_1^--y_2^-)}e^{ix_3p^+y_2^-} H_{\mu\nu}\left(\{x_i\},p,q,\ell,\ell_h,k_{2T},k_{3T},\{y_i\}\right).
\label{hard-{xi}}
\eea

\subsection{Quark-gluon double scattering}
In this subsection we calculate the double-scattering contribution for the virtual-photon-quark ($\gamma^*+q$) interaction channel, as illustrated in Fig.~\ref{fig-DIS}(b), which involves a quark and a gluon in the initial state. They will be referred to as quark-gluon double scattering, in which there is first a hard photon-quark scattering, and then the produced parton undergoes a second scattering with another initial gluon from the nucleus. To simplify our discussion, we classify the secondary scattering as ``soft'' or ``hard'' \cite{Wang:2001ifa,Guo:2000nz,Guo:1997it,Zhang:2003yn,Zhang:2003wk,Zhang:2004qm},  depending on whether the exchanged gluon momentum [either $k_g$ or $k_g'$ in Fig.~\ref{fig-DIS}(b)] becomes zero or remains finite,  respectively, when $k_T\to 0$. The final amplitudes of the cut diagrams contain ``soft'', or ``hard'' contributions and their interferences, often referred to as soft-soft, hard-hard, soft-hard and hard-soft contributions. We will first study the central-cut diagrams, which represent the classical double-scattering picture; then we compute the virtual contributions, and finally we come back to the asymmetric-cut diagrams, which represent the interference between single- and triple-scattering processes. As we will show below, both central-cut diagrams and virtual contributions contain divergences, while the sum of all the asymmetric-cut diagrams is free of any divergence, and only contributes to the NLO finite terms.

\subsubsection{Central cut (real corrections)}
For real corrections, there are in total 16 diagrams corresponding to four different kinds of subprocesses mentioned above:  soft-soft double scattering, hard-hard double scattering and the interferences between them, as shown in Fig.~\ref{fig-central}. Let us take the soft-soft double scattering in Fig.~\ref{fig-central}(a) as an example to outline the essential steps for calculating the NLO contributions to the transverse momentum broadening, and all the other subprocesses could be evaluated in the same manner. 
\bef
\psfig{file=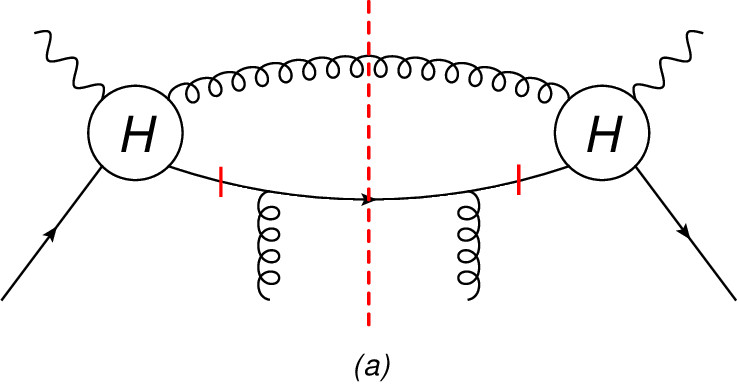, width=1.5in}
\hskip 0.3in
\psfig{file=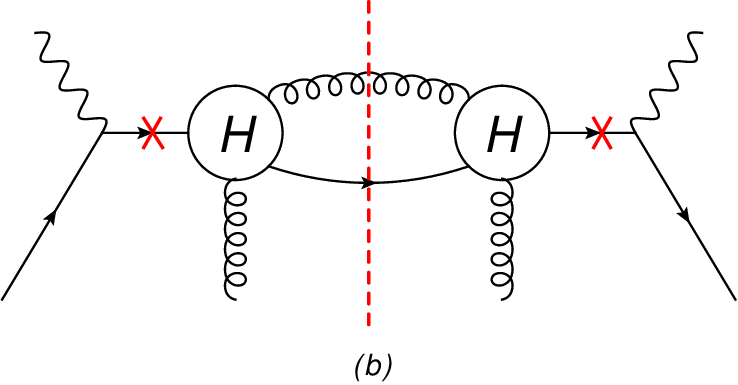, width=1.5in}
\hskip 0.3in
\psfig{file=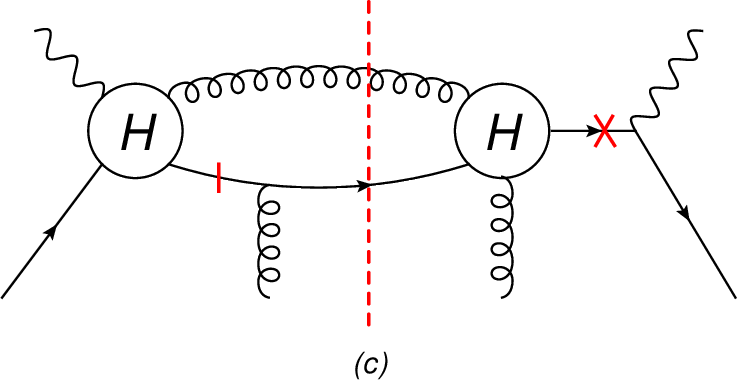, width=1.5in}
\hskip 0.3in
\psfig{file=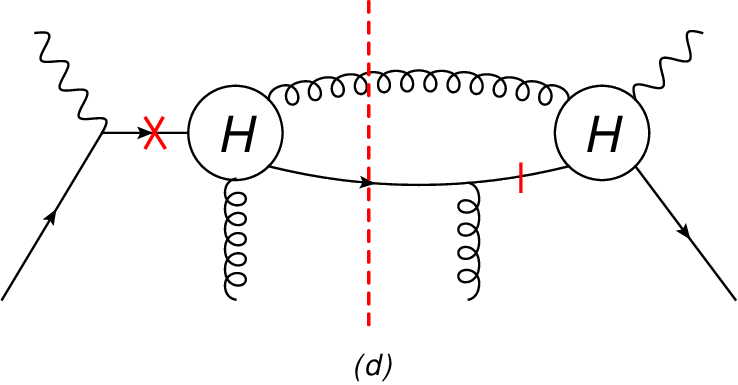, width=1.5in}
\caption{The central-cut diagrams for (a) soft-soft, (b) hard-hard, (c) soft-hard, and (d) hard-soft double scatterings in SIDIS. The short bars indicate the propagators where the soft poles arise, while the crosses indicate the propagators where the hard poles arise. The ``$H$" blobs represent the hard $2\to 2$ processes as shown in Fig. \ref{fig-Hrepresent}.}
\label{fig-central}
\eef

\bef
\psfig{file=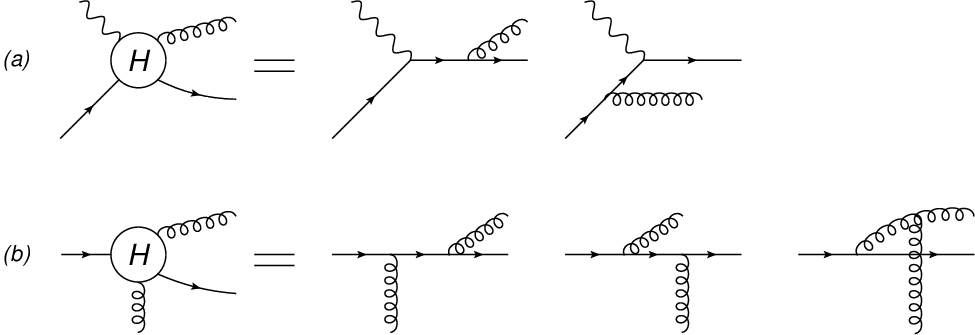, width=4in}
\caption{The representations of hard $2\to 2$ processes for (a) photon-quark interaction, and (b) quark-gluon interaction. }
\label{fig-Hrepresent}
\eef

To perform the collinear expansion in Eq.~\eqref{eq-HTgeneral}, we first integrate out the parton momentum fractions $x_1$, $x_2$, and $x_3$ with the help of either the contour integration or the kinematic $\delta$ function in the final-state phase space, and then perform the $k_T$ expansion directly. Starting from the high-twist general formalism as shown in Eq. (\ref{eq-HTgeneral}), and following the original setup of high-twist expansion as developed by Qiu and Sterman \cite{Luo:1992fz,Luo:1994np}, we set $k_{2T}=k_{3T}=k_T$. This is the most convenient way to perform a twist-4 calculation for soft-soft double scattering. A detailed explanation on how to choose the correct setup of $k_T$ flow to ensure gauge invariance was provided in Ref. \cite{Kang:2016ron}. Within this setup, the $\ell_{hT}^2$-weighted hadronic tensor for soft-soft double scattering can then be written as,
\bea
\frac{d\langle \ell_{hT}^2W_{\mu\nu}^D\rangle}{dz_h}=&\sum_qe_q^2\int\frac{dz}{z}D_{h/q}(z)  \ell_{hT}^2
 \left[-\frac{g^{\alpha\beta}}{2(1-\epsilon)}\right]\frac{1}{2}\frac{\partial^2}{\partial k_T^{\alpha}\partial k_T^{\beta}}
\left[
\int dx_1 dx_2 dx_3 T_{qg}(\{x_i\})
H_{\mu\nu}(\{x_i\},p,q,\ell,\ell_h,k_T)\right]_{k_T=0}.
\eea
The two propagators which will be used to perform the contour integrals are marked by short bars in Fig.~\ref{fig-central}(a) and can be expressed as follows:
\bea
\frac{1}{(\ell-x_2p-k_T)^2+i\epsilon} &=\frac{x}{{\hat u}}\frac{1}{x_2-x_D-i\epsilon},
\label{also-prop}
\\
\frac{1}{\left[\ell-(x_2-x_3)p-k_T\right]^2-i\epsilon}&=\frac{x}{{\hat u}}\frac{1}{x_2-x_3-x_D+i\epsilon}.
\label{propagators}
\eea
On the other hand, the two-body final-state phase space integral at central-cut is given by
\bea
dPS^{(C)}=\frac{1}{8\pi}\left(\frac{4\pi}{Q^2}\right)^{\epsilon}
\frac{1}{\Gamma(1-\epsilon)}\int dx\,\delta\left(x_1+x_2-x-x_C\right)\hat{z}^{-\epsilon}(1-\hat{z})^{-\epsilon}\hat{x}^{\epsilon}(1-\hat{x})^{-\epsilon},
\label{delta}
\eea
where the $\delta$ function $\delta\left(x_1+x_2-x-x_C\right)$ comes from the on-shell condition for the unobserved final-state gluon. Here the momentum fractions $x$, $x_C$, and $x_D$ in Eqs.~\eqref{also-prop}, \eqref{propagators} and \eqref{delta} are given by
\bea
x=\frac{Q^2+2q\cdot \ell}{2p\cdot (q-\ell)},
\qquad
x_C=x\frac{k_T^2-2\ell\cdot k_T}{{\hat t}},
\qquad
x_D=x\frac{2\ell\cdot k_T-k_T^2}{\hat u}.
\label{eq-xCD}
\eea
Now we are able to integrate over $\{x_i\}$,
\bea
&\int dx_1dx_2dx_3e^{ix_1p^+y^-}e^{ix_2p^+(y_1^--y_2^-)}e^{ix_3p^+y_2^-}
\frac{1}{x_2-x_D-i\epsilon}\frac{1}{x_2-x_3-x_D+i\epsilon}\delta\left(x_1+x_2-x-x_C\right)
\nnu
& = e^{i(x+x_C-x_D)p^+y^-}e^{ix_Dp^+(y_1^--y_2^-)}(2\pi)^2\theta(y_2^-)\theta(y_1^--y^-).
\eea
In the above equation, two of the integrations over $\{x_i\}$ are carried out by contour integrations, which lead to the $\theta$ functions, indicating the order of the two scatterings. The third integration over $\{x_i\}$ is fixed by the $\delta$ function from the final-state phase space. After the integration, the parton momentum fractions  $\{x_i\}$ are fixed as follows:
\bea
x_1=x+x_C-x_D,
\qquad
x_2=x_D,
\qquad
x_3=0.
\label{eq-softxi}
\eea
As we can see here, in the collinear limit $k_T\to 0$, $x_C=x_D=0$ according to Eq. (\ref{eq-xCD}). Thus the momentum fraction for the initial quark is finite $x_1=x$, while the momentum fractions for the initial gluons on both sides of the cut line become zero ($k_g\to 0$ and $k_g'\to 0$). This is why we refer to this process as soft-soft double scattering.

The next critical step, which is the key point in the high-twist calculation, is to perform the collinear expansion. With the help of the identity \cite{Kang:2013ufa},
\bea
\frac{\partial^2\big[T(\{x_i\}) H_{\mu\nu}(\{x_i\}, k_T)\big]}{\partial k_T^\alpha \partial k_T^\beta} 
=
\frac{\partial^2 T}{\partial x_i \partial x_j} 
\left[\frac{\partial x_i}{\partial k_T^\alpha} \frac{\partial x_j}{\partial k_T^\beta} H_{\mu\nu} \right] 
+ \frac{\partial T}{\partial x_i} 
\left[\frac{\partial^2 x_i}{\partial k_T^\alpha \partial k_T^\beta} H_{\mu\nu}
+ \frac{\partial x_i}{\partial k_T^\alpha}\frac{\partial H_{\mu\nu}}{\partial k_T^\beta} 
+\frac{\partial x_i}{\partial k_T^\beta} \frac{\partial H_{\mu\nu}}{\partial k_T^\alpha}\right]
+T \frac{\partial^2 H_{\mu\nu}}{\partial k_T^\alpha \partial k_T^\beta},
\label{eq-identity}
\eea
where repeated indices imply summations, we substitute the parton momentum fractions $\{x_i\}$ in 
Eq.~\eqref{eq-softxi}, and then carry out the collinear expansion of the hard part. At the end of the day, we have
\bea
\frac{d\langle\ell_{hT}^2W^D\rangle^{ss}_C}{dz_h}=&\frac{2\alpha_s}{N_c}z_h^2(2\pi)^3(1-\epsilon)\frac{\alpha_s}{2\pi}\int\frac{dx}{x}\int\frac{dz}{z}D_{h/q}(z)\left(\frac{4\pi\mu^2}{Q^2}\right)^{\epsilon}
\frac{1}{\Gamma(1-\epsilon)}\hat{z}^{-\epsilon}(1-\hat{z})^{-\epsilon}\hat{x}^{\epsilon}(1-\hat{x})^{-\epsilon}\nnu
&\times\left[x^2\frac{d^2}{dx^2}T_{qg}(x,0,0)D_{2}^{ss}+x\frac{d}{dx}T_{qg}(x,0,0)D_{1}^{ss}+T_{qg}(x,0,0)D_{0}^{ss}\right].
\label{eq-ssgeneral}
\eea
Here and throughout the later part of this paper, $W^D$ (H) stands for the combination of metric and longitudinal contributions by contracting $W^D_{\mu\nu}$ ($H_{\mu\nu}$) with $-g^{\mu\nu}$ and $p^{\mu}p^{\nu}$ separately. In Eq. (\ref{eq-ssgeneral}), $\mu$ is the mass scale introduced to keep the coupling constant dimensionless $g\to g\mu^{\epsilon}$, and the superscript ``$ss$'' represents the soft-soft contributions. There are three terms in Eq.~\eqref{eq-ssgeneral}: the first two are the derivative terms, and the third one is the nonderivative term, and they are related to the hard part coefficient function $H(\{x_i\}, k_T)$ as
\bea
D_{2}^{ss}=&\frac{1}{2\hat z^2}\left(\frac{1}{\hat t}+\frac{1}{\hat u}\right)^2\frac{\ell_T^4}{(1-\epsilon)^2}H,
\\
D_{1}^{ss}=&-\frac{1}{2\hat z^2}\left(H+
\frac{\ell_T^2}{1-\epsilon}\frac{\partial H}{\partial y_1}\right)\frac{\ell_T^2}{1-\epsilon},
\\
D_{0}^{ss}=&\frac{1}{2\hat z^2}\left(\frac{1}{4}\frac{\ell_T^2}{1-\epsilon}\frac{\partial^2H}{\partial y_1^2}
-\frac{\partial H}{\partial y_2}\right)\frac{\ell_T^2}{1-\epsilon},
\eea 
where $y_1=\ell\cdot k_T$, $y_2=k_T^2$ and the arguments in $H$ are suppressed. In arriving at Eq.~\eqref{eq-ssgeneral} from Eq.~\eqref{eq-identity}, one realizes that only derivatives with respect to $x_1$ contribute to the final result; thus we change the partial derivative with respect to $x_1$ into the form of full derivative with respect to $x$ (recall $x_1\to x$ at $k_T\to 0$). The first derivative with respect to $x_2$ will generate $(y_1^- - y_2^-)$, and thus when combined with the matrix element, it vanishes due to the fact that the gluon field strengths commute on the light cone, as explained clearly in Refs. \cite{Luo:1992fz,Luo:1994np}. The second derivative with respect to $x_2$ gives rise to a ``contact'' term when combined with the corresponding asymmetric-cut diagrams. The ``contact'' terms generally do not have nuclear size enhancement and thus we neglect them in our study; see the explanation in Ref.~\cite{Kang:2013ufa} and also discussions in Sec.~\ref{sec-asym} below. Finally since $x_3=0$ is independent of $k_T$, no expansion over $x_3$ is needed. 

In the above equations, the hard part coefficients $D_{i}^{ss}(i=2,1,0)$ are functions of parton Mandelstam variables $\hat s, ~\hat t$ and $\hat u$, which can be expressed in terms of $Q^2$, $\hat x$ and $\hat z$ as
\bea
\hat s=\frac{1-\hat x}{\hat x}Q^2,
\qquad
\hat t= - \frac{1-\hat z}{\hat x}Q^2,
\qquad
\hat u= - \frac{\hat z}{\hat x}Q^2.
\eea
Thus we see that the integrals over $\hat x$ and $\hat z$ will contain divergences when $\hat x\to1$ and $\hat z \to 1$. Note that we do not have to worry about the divergences when $\hat x\to 0$ and $\hat z\to 0$ since they are outside the physical regions ($\hat x > x_B$ and $\hat z > z_h$). The main task now is to isolate all the divergences, and combine them accordingly. Let us define the following common factor 
\bea
I=\hz^{-\epsilon} (1-\hz)^{-\epsilon} \hx^\epsilon (1-\hx)^{-\epsilon},
\eea
which will be used repeatedly below. To perform the $\epsilon$ expansion for the hard part coefficients $I\times D_{i}^{ss}$, we use the following formulas \cite{Altarelli:1979ub}:
\bea
\hz^{-\epsilon} (1-\hz)^{-\epsilon-1} =& -\frac{1}{\epsilon} \delta(1-\hz)+\frac{1}{(1-\hz)_+} 
- \epsilon \left(\frac{\ln(1-\hz}{1-\hz}\right)_+ - \epsilon \frac{\ln\hz}{1-\hz} 
+ {\cal O}(\epsilon^2),
\\
\hx^{\epsilon} (1-\hx)^{-\epsilon-1} =& -\frac{1}{\epsilon} \delta(1-\hx)+\frac{1}{(1-\hx)_+} 
- \epsilon \left(\frac{\ln(1-\hx}{1-\hx}\right)_+ + \epsilon \frac{\ln\hx}{1-\hx}
+ {\cal O}(\epsilon^2),
\\
\hz^{-\epsilon} (1-\hz)^{-\epsilon} =& 1-\epsilon \ln\hz - \epsilon\ln(1-\hz)
+ {\cal O}(\epsilon^2),
\\
\hx^{\epsilon} (1-\hx)^{-\epsilon} =& 1+\epsilon \ln\hx - \epsilon \ln(1-\hx)
+ {\cal O}(\epsilon^2),
\eea
where the usual ``plus" function is defined as
\bea
\int_{0}^1dz\frac{f(z)}{(1-z)_+} \equiv \int_{0}^1dz\frac{f(z)-f(1)}{1-z}.
\eea
Finally we have
\bea
I\times D_{2}^{ss}=&-\frac{1}{\epsilon}C_F\delta(1-\hat z)(1-\hat x)(1+\hat x^2)+\cdots,
\label{D2}
\\
I\times D_{1}^{ss}=&-\frac{1}{\epsilon}C_F\delta(1-\hat z)(4\hat x^3-5\hat x^2-1)+\cdots,
\label{D1}
\\
I\times D_{0}^{ss}=&C_F\bigg[\frac{2}{\epsilon^2}\delta(1-\hat z)\delta(1-\hat x)+\frac{4}{\epsilon}\delta(1-\hat z)\delta(1-\hat x)
-\frac{1}{\epsilon}\delta(1-\hat x)\frac{1+\hat z^2}{\hat z^2(1-\hat z)_+}
\nnu
&
-\frac{1}{\epsilon}\delta(1-\hat z)\frac{1+\hat x^2(6\hat x^2-14\hat x+9)}{(1-\hat x)_+}\bigg]+\cdots,
\label{D0}
\eea
where the ellipses denote finite contributions. It is instructive to point out that all the divergent terms above come from the metric contribution, not from the longitudinal contribution. However, the longitudinal part does contribute to finite terms. This feature holds true in all the other processes as well. For the divergent pieces associated with derivative terms in the above expression, we further perform integration by parts to convert them into the form of nonderivative terms \cite{Kang:2012ns,VITEV:2014wza}. We have the final divergent piece in soft-soft double scattering as
\bea
C_F\int_{x_B}^1\frac{dx}{x}T_{qg}(x,0,0)\left[\frac{2}{\epsilon^2}\delta(1-\hat x)\delta(1-\hat z)-\frac{1}{\epsilon}\delta(1-\hat x)\frac{1+\hat z^2}{\hat z^2(1-\hat z)_+}-\frac{1}{\epsilon}\delta(1-\hat z)\frac{1+\hat x^2}{(1-\hat x)_+}\right],
\eea 
where we have used the boundary condition $T_{qg}(x,0,0)=0$ when $x\to 1$ in preforming the integration by parts, which is valid under the approximation of neglecting the Fermi motion of a nucleon inside a nucleus. From the divergent piece, we can see that soft-soft double scattering contains both soft-collinear and collinear divergences, which are identified as double-pole $1/\epsilon^2$ and single-pole $1/\epsilon$, respectively. 
On the other hand, the finite terms associated with derivative and nonderivative terms [as denoted by the ellipses in Eqs.~\eqref{D2}, \eqref{D1}, and \eqref{D0}] are combined into a single term denoted as $H_{qg-C}^{ss}\otimes T_{qg}$, with the expression given by Eq.~\eqref{eq-finite-ss} in the Appendix.

Likewise, we can also compute the diagrams of hard-hard double scattering as shown in Fig.~\ref{fig-central}(b), where the radiated gluon is induced by the secondary quark-nucleus scattering, following the first quark-photon interaction. In this process, one can either use the original Qiu-Sterman setup ($k_{2T}=k_{3T}=k_T$), or apply the one shown in Fig. \ref{fig-DIS}(b), which is more clear to demonstrate gauge invariance. We have checked that these two setups lead to exactly the same result. To simplify our presentation, we use the same scenario as that in soft-soft double scattering (Qiu-Sterman setup).  In this process, it is straightforward to show that the exchanged gluon momenta ($k_g$ and $k_g'$ ) remain finite in the collinear limit $k_T\to 0$, and thus it is referred to as a hard scattering. Specifically the two propagators marked by the crosses have the following expressions:
\bea
\frac{1}{(x_1 p+ q)^2+i\epsilon} &=\frac{x_B}{Q^2} \frac{1}{x_1 - x_B + i\epsilon},
\\
\frac{1}{\left[(x_1+x_3) p+ q\right]^2+i\epsilon} &=\frac{x_B}{Q^2} \frac{1}{x_1 + x_3 - x_B + i\epsilon}.
\eea
At the same time, the on-shell condition for the unobserved gluon leads to 
\bea
\delta\left[\left((x_1+x_2)p+k_T+q - \ell\right)^2\right] \to \delta(x_1+x_2-x-x_C).
\eea
Thus the contour integrals and the above kinematic $\delta$ function fix $\{x_i\}$ as
\bea
x_1 = x_B,
\qquad
x_2 = x+x_C- x_B,
\qquad
x_3 = 0,
\eea
from which we find that the gluon momenta associated with the second scattering remain finite when $k_T\to 0$ as
\bea
k_g\to (x-x_B) p,
\qquad
k_g' \to (x-x_B) p,
\eea
hence the name hard-hard double scattering.

Following the same steps as we have outlined in soft-soft double scattering, we can write down the contributions from hard-hard double scattering, where only nonderivative term contributes to the final result:
\bea
\frac{d\langle\ell_{hT}^2W^D\rangle^{hh}_C}{dz_h}=&\frac{\alpha_s^2}{N_c}z_h^2(2\pi)^3(1-\epsilon)\int\frac{dx}{x}\int\frac{dz}{z}D_{h/q}(z)\left(\frac{4\pi\mu^2}{Q^2}\right)^{\epsilon}
\frac{1}{\Gamma(1-\epsilon)}
\hz^{-\epsilon} (1-\hz)^{-\epsilon} \hx^\epsilon (1-\hx)^{-\epsilon}
\nnu
&\times
T_{qg}(x_B,x-x_B,0)D_0^{hh},
\eea
where the superscript ``$hh$'' represents the hard-hard scattering contribution. Perform the $\epsilon$ expansion, we have 
\bea
I\times D_0^{hh}=&C_A\left[\frac{2}{\epsilon^2}\delta(1-\hat z)\delta(1-\hat x)
-\frac{2}{\epsilon}\delta(1-\hat x)\frac{1+\hat z^2}{(1-\hat z)_+}\frac{C_F/C_A(1-\hat z)^2+\hat z}{\hat z^2}
-\frac{2}{\epsilon}\delta(1-\hat z)\frac{1}{(1-\hat x)_+}+\cdots\right].
\label{hard-hard}
\eea
The finite term denoted by the ellipsis comes from the metric part only, with the explicit expression $H_{qg-C}^{hh}\otimes T_{qg}$ given in Eq.~\eqref{eq-finite-hh} in the Appendix.
 
Finally let us turn to the interference diagrams between soft and hard scatterings. The calculation is similar, but one has to be very careful when choosing the correct setup for the $k_T$-flow to ensure the gauge invariance of the final result; a detailed discussion of this point can be found in Ref. \cite{Kang:2016ron}.  For the soft-hard scattering contributions as shown in Fig.~\ref{fig-central}(c), we choose the setup as shown in Fig. \ref{fig-DIS}(b), and perform the collinear expansion as in Eq.~\eqref{eq-HTgeneral}. The result for soft-hard double scattering can be written as
\bea
\frac{d\langle\ell_{hT}^2W^D\rangle^{sh}_C}{dz_h}=&\frac{2\alpha_s}{N_c}z_h^2(2\pi)^3(1-\epsilon)
\frac{\alpha_s}{2\pi}\int\frac{dx}{x}\int\frac{dz}{z}D_{h/q}(z)\left(\frac{4\pi\mu^2}{Q^2}\right)^{\epsilon}
\frac{1}{\Gamma(1-\epsilon)}\hat{z}^{-\epsilon}(1-\hat{z})^{-\epsilon}\hat{x}^{\epsilon}(1-\hat{x})^{-\epsilon}
\nnu
&\times
\left[x\frac{d}{dx}T_{qg}(x,0,x_B-x)D_1^{sh}
+x\left.\frac{d}{dx_2}T_{qg}(x,x_2,x_B-x)\right|_{x_2\to 0}D_{12}^{sh}
+T_{qg}(x,0,x_B-x)D_0^{sh}\right].
\eea
Again, the divergences in each term of the above equation can be identified as follows:
\bea
I\times D_1^{sh}=&-\frac{1}{\epsilon}\frac{C_A}{2}\delta(1-\hat z)(1+\hat x)
\cdots ,
\\
I\times D_{12}^{sh}=&\cdots ,
\\
I\times D_0^{sh}=&\frac{C_A}{2}\bigg\{-\frac{2}{\epsilon^2}\delta(1-\hat z)\delta(1-\hat x)-\frac{4}{\epsilon}\delta(1-\hat z)\delta(1-\hat x)
+\frac{1}{\epsilon}\delta(1-\hat x)\frac{1+\hat z^2}{\hat z^2(1-\hat z)_+}[\hat z+2C_F/C_A(1-\hat z)]
\nnu
&
\hspace{26pt}
+\frac{1}{\epsilon}\delta(1-\hat z)\frac{1+2\hat x-\hat x^2}{(1-\hat x)_+}+\cdots
\bigg\}.
\eea
Similarly to the soft-soft double scattering, performing partial integration to convert the derivative of the quark-gluon correlation function $T_{qg}$ to  $T_{qg}$ itself leads to the divergent part
\bea
C_A\int_{x_B}^1\frac{dx}{x}T_{qg}(x,0,x_B-x)
\bigg\{&-\frac{1}{\epsilon^2}\delta(1-\hat x)\delta(1-\hat z)
+\frac{1}{\epsilon}\delta(1-\hat x)\frac{1+\hat z^2}{\hat z^2(1-\hat z)_+}\left[\frac{\hat z}{2}+\frac{C_F}{C_A}(1-\hat z)\right]
+\frac{1}{\epsilon}\delta(1-\hat z)\frac{1+\hat x}{2(1-\hat x)_+}
\nnu
&-\frac{1}{\epsilon}\delta(1-\hat x)\delta(1-\hat z)
\bigg\},
\label{soft-hard}
\eea 
while the finite contribution denoted as $H_{qg-C}^{sh}\otimes T_{qg}$ is given by Eq.~\eqref{eq-finite-sh} in the Appendix. Like in hard-hard double scattering, the finite contribution in soft-hard double scattering comes from the metric contribution only, and the longitudinal part does not contribute. This also holds true for the hard-soft double-scattering process.

The process of hard-soft double scattering as shown in Fig.~\ref{fig-central}(d) is simply the complex conjugate of the soft-hard double scattering, its contribution can be easily obtained by replacing the matrix element in soft-hard process as follows
\bea
T_{qg}(x,0,x_B-x)\to T_{qg}(x_B,x-x_B,x-x_B).
\eea
Therefore, the divergent part in this process is
\bea
&C_A\int_{x_B}^1\frac{dx}{x}T_{qg}(x_B,x-x_B,x-x_B)\left\{-\frac{1}{\epsilon^2}\delta(1-\hat x)\delta(1-\hat z)
+\frac{1}{\epsilon}\delta(1-\hat x)\frac{1+\hat z^2}{\hat z^2(1-\hat z)_+}\left[\frac{\hat z}{2}+\frac{C_F}{C_A}(1-\hat z)\right]
\right.
\nnu
&
\hspace{160pt}\left.+\frac{1}{\epsilon}\delta(1-\hat z)\frac{1+\hat x}{2(1-\hat x)_+}
-\frac{1}{\epsilon}\delta(1-\hat x)\delta(1-\hat z)
\right\},
\label{hard-soft}
\eea 
and the finite part denoted as $H_{qg-C}^{hs}\otimes T_{qg}$ can be found in Eq.~\eqref{eq-finite-hs} in the Appendix.

Combining all the results from soft-soft, hard-hard, soft-hard and hard-soft contributions, we obtain the result for real corrections from central-cut diagrams, 
\bea
\frac{d\langle\ell_{hT}^2\sigma^D\rangle^{\rm (C)}}{d{\cal PS}}=&\sigma_h\frac{\alpha_s}{2\pi}\sum_qe_q^2\int\frac{dx}{x}\int\frac{dz}{z}D_{h/q}(z)\left(\frac{4\pi\mu^2}{Q^2}\right)^{\epsilon}
\frac{1}{\Gamma(1-\epsilon)}
\Bigg\{\frac{2}{\epsilon^2}C_F\delta(1-\hat x)\delta(1-\hat z)T_{qg}(x,0,0)
-\frac{1}{\epsilon}\delta(1-\hat x)
\nnu
&
\times C_F\frac{1+\hat z^2}{(1-\hat z)_+}T_{qg}(x,0,0)
-\frac{1}{\epsilon}\delta(1-\hat z)\bigg[
C_F\frac{1+\hat x^2}{(1-\hat x)_+}T_{qg}(x,0,0)
+C_A\frac{2}{(1-\hat x)_+}T_{qg}(x_B,x-x_B,0)
\nnu
&
-\frac{C_A}{2}\frac{1+\hat x}{(1-\hat x)_+}\big(T_{qg}(x,0,x_B-x)+T_{qg}(x_B,x-x_B,x-x_B)\big)\bigg]
-\frac{2}{\epsilon}\delta(1-\hat x)\delta(1-\hat z)T_{qg}(x,0,0)
\nnu
&
+H_{qg}^{C-R}\otimes T_{qg}\Bigg\},
\label{eq-cfinal}
\eea
where the finite contribution $H_{qg}^{C-R}\otimes T_{qg}$ has the following form
\bea
H_{qg}^{C-R}\otimes T_{qg} = H_{qg-C}^{ss}\otimes T_{qg} + H_{qg-C}^{hh}\otimes T_{qg} 
+ H_{qg-C}^{sh}\otimes T_{qg} + H_{qg-C}^{hs}\otimes T_{qg},
\label{finite-C-R}
\eea
with all the terms on the right-hand side given in Eqs.~\eqref{eq-finite-ss}, \eqref{eq-finite-hh}, \eqref{eq-finite-sh}, and \eqref{eq-finite-hs}, respectively. It is instructive to point out that even though hard-hard double scattering, soft-hard and hard-soft scattering all have double-pole $1/\epsilon^2$ terms $\propto C_A$, they cancel between them, and thus the remaining $1/\epsilon^2$ terms entirely come from the soft-soft double-scattering contribution, which has a color factor $C_F$, and is exactly opposite to those in the virtual corrections as we will show in the next subsection.

\subsubsection{Virtual corrections}
In this subsection, we calculate the virtual corrections in quark-gluon double scattering, which have to be included to ensure unitarity and infrared safety of the final result. The relevant generic Feynman diagrams are shown in Fig.~\ref{fig-virtual}, in which the blob is given by Fig.~\ref{fig-vertex}. The incoming parton momenta involved in the double scatterings follow the same convention as those in Fig.~\ref{fig-DIS}(b), or the LO diagram shown in Fig.~\ref{fig-LO}. In this case, it is important to realize that all the asymmetric-cut diagrams give no contribution to the $\ell_{hT}^2$-weighted differential cross section. This is because the kinematic $\delta$ function $\delta^{n-2}(\ell_{hT})$ from final-state phase space leads to $\int d^{n-2} \ell_{hT} \ell_{hT}^2 \delta^{n-2}(\ell_{hT})=0$. Thus we only have to consider the central-cut diagrams. 
\bef
\psfig{file=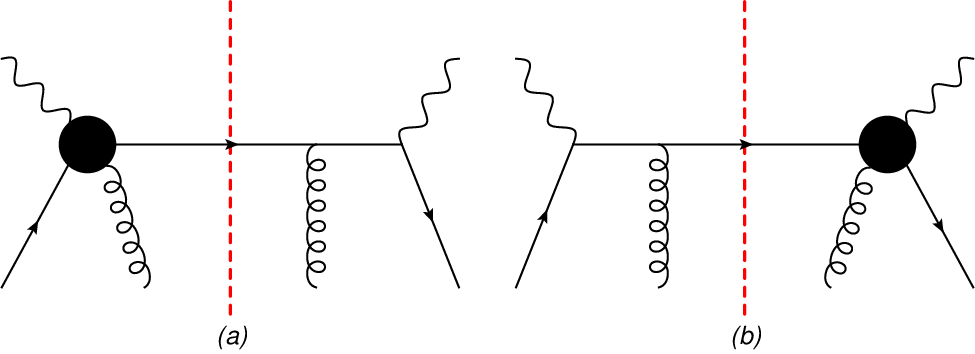, width=3in}
\caption{The virtual diagrams in the calculation of transverse momentum broadening at NLO in SIDIS. The incoming parton momenta involved in the double scatterings follow the same convention as in Fig.~\ref{fig-DIS}(b), or the LO diagram shown in Fig.~\ref{fig-LO}.}
\label{fig-virtual}
\eef
\bef
\psfig{file=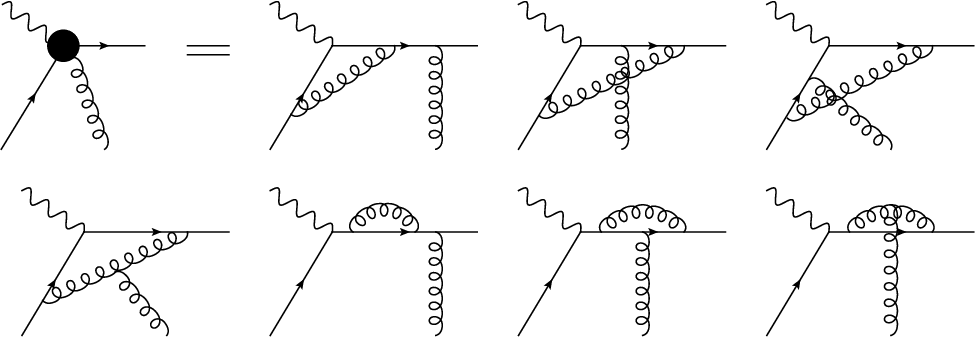, width=5in}
\caption{One-loop corrections to the quark-photon-quark vertex with gluon attachment, corresponding to the blob in Fig.~\ref{fig-virtual}.}
\label{fig-vertex}
\eef

Two diagrams in Fig.~\ref{fig-virtual} are simply complex conjugates of each other, so they should have the same result. The actual calculation is quite involved and tedious, and it contains a significant amount of tensor reductions and integrations. Nevertheless, the calculation is straightforward. The results can be decomposed into two types of color factors, $C_F$ and $C_A$, and it turns out that terms associated with $C_A$ cancel out and only terms with the color $C_F$ remain. The final result for the virtual correction is quite simple and has exactly the same structure as the virtual correction at leading twist,
\bea
\frac{d\langle\ell_{hT}^2\sigma^D\rangle}{d{\cal PS}}^{\rm (V)}=&\sigma_h
\frac{\alpha_s}{2\pi}\int\frac{dx}{x}T_{qg}(x,0,0)\int\frac{dz}{z}D_{h/q}(z)\delta(1-{\hat x})\delta(1-{\hat z})
\left(\frac{4\pi\mu^2}{Q^2}\right)^{\epsilon}
\frac{1}{\Gamma(1-\epsilon)}C_F\left(-\frac{2}{\epsilon^2}-\frac{3}{\epsilon}-8\right).
\label{eq-virtual}
\eea
A similar structure also appears in the virtual correction to the transverse-momentum-weighted spin-dependent cross section at the twist 3 \cite{Kang:2012ns,VITEV:2014wza, Vogelsang:2009pj}. It is important to note that the soft-collinear divergence ($1/\epsilon^2$ term) in the above virtual corrections should cancel that in the real diagrams in order to establish the NLO collinear factorization at twist 4. We will check this cancellation when we combine the results of all the diagrams together. For later convenience, we write out the finite term in the virtual contribution:  
\bea
H_{qg}^{C-V}\otimes T_{qg} = -8C_F\delta(1-{\hat x})\delta(1-{\hat z})T_{qg}(x,0,0).
\label{finite-virtual}
\eea

\subsubsection {Asymmetric cut}
\label{sec-asym}
We now turn to the asymmetric-cut diagrams, which represent the interferences between single and triple scatterings. They include both left-cut and right-cut diagrams as shown in Figs.~\ref{fig-left} and \ref{fig-right}, respectively. Since two additional scattered gluons are always on the same side, there will be no hard-hard scattering contributions. Thus there are only three different kinds of subprocesses for asymmetric-cut diagrams: soft-soft, soft-hard and hard-soft rescatterings.
\bef
\psfig{file=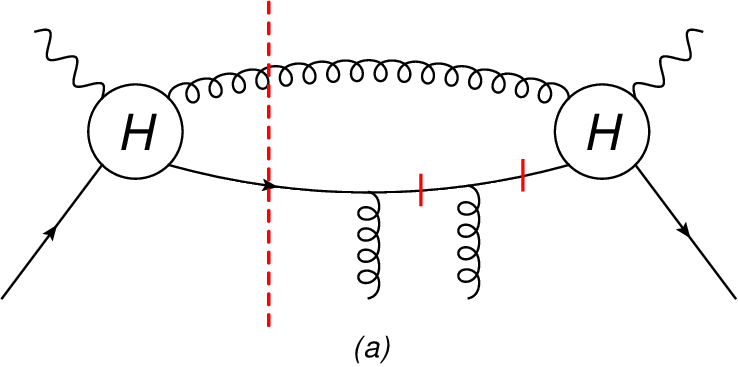, width=2in}
\hskip 0.3in
\psfig{file=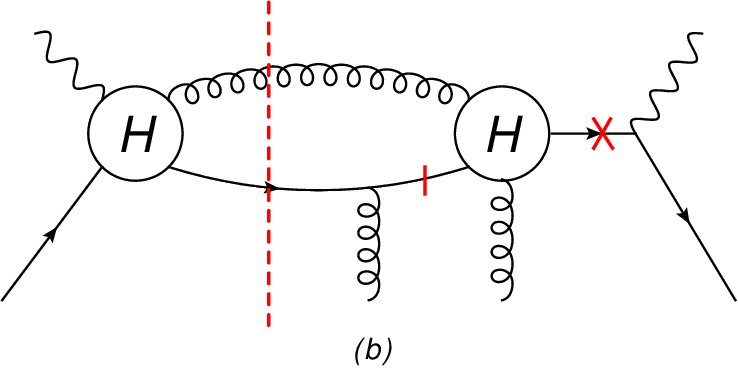, width=2in}
\hskip 0.3in
\psfig{file=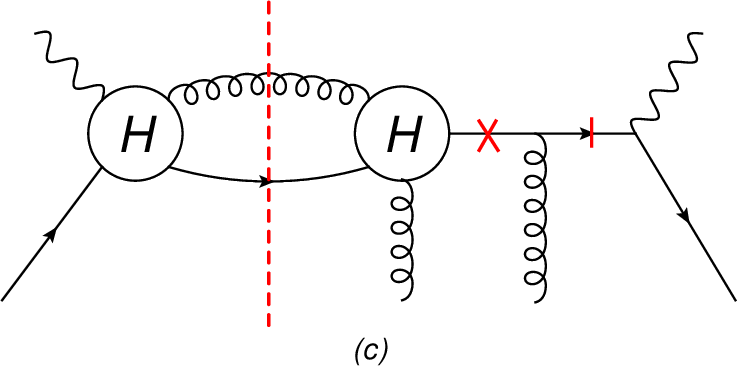, width=2in}
\caption{The left-cut diagrams for (a) soft-soft, (b) soft-hard, and (c) hard-soft rescattering processes in SIDIS. The short bars (crosses) indicate the propagators where the soft (hard) poles arise.}
\label{fig-left}
\eef
\bef
\psfig{file=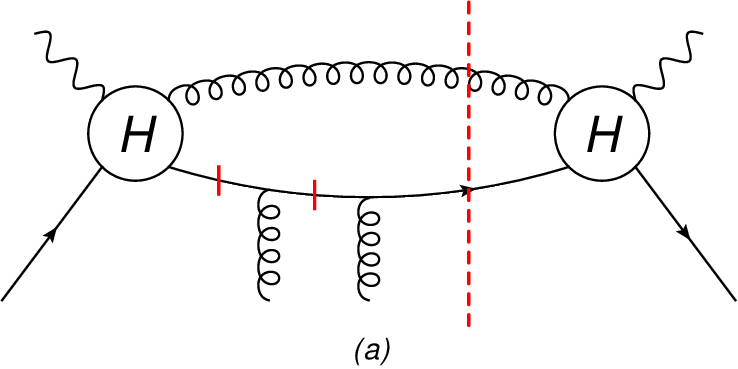, width=2in}
\hskip 0.3in
\psfig{file=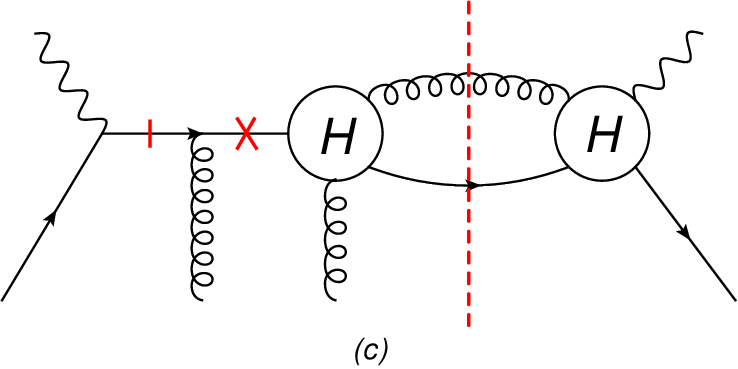, width=2in}
\hskip 0.3in
\psfig{file=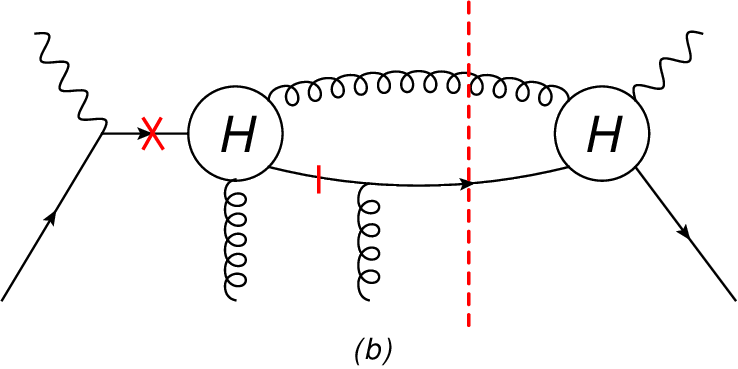, width=2in}
\caption{The right-cut diagrams for (a) soft-soft, (b) soft-hard, and (c) hard-soft rescattering processes in SIDIS. The short bars (crosses) indicate the propagators where the soft (hard) poles arise.}
\label{fig-right}
\eef

The soft-soft rescatterings of single-triple interference are shown in Figs. \ref{fig-left}(a) and \ref{fig-right}(a). Let us take Fig.~\ref{fig-left}(a) as an example, in which the relevant propagators (marked by the short bars) are,
\bea
\frac{1}{(\ell+x_2p+k_T)^2 - i\epsilon} &= -\frac{x}{\hat u} \frac{1}{x_2 - x_E - i\epsilon},
\\
\frac{1}{(\ell+x_3 p)^2- i\epsilon} & = -\frac{x}{\hat u} \frac{1}{x_3 - i\epsilon}.
\eea
Together with the on-shell condition for the unobserved gluon, which gives $\delta(x_1-x)$, we have 
\bea
x_1 = x, 
\qquad
x_2 = x_E,
\qquad
x_3 = 0,
\eea
where $x_E$ is given by
\bea
x_E = \frac{x}{\hat u} \left(2\ell\cdot k_T + k_T^2 \right).
\eea
Only $x_2$ which is set to $x_E$ by the pole in the first propagator depends on $k_T$ and  vanishes when $k_T\rightarrow 0$. We further find that the hard part coefficient $H(\{x_i\}, k_T)$ is independent of $k_T$, and therefore according to Eq.~\eqref{eq-identity}, all the terms associated with the derivative of the hard part coefficient vanish and only the derivative terms with respect to $x_2$ survive. 
As we have pointed out already when discussing soft-soft double scattering in central-cut diagrams, the single-derivative term with respect to $x_2$ vanishes due to the commutation of gluon field strengths on the light cone. 
We further find that the double-derivative term w.r.t. $x_2$ leads to the ``contact'' contribution to the final result. For example, when we combine the soft-soft contributions in central-cut, left-cut and right-cut diagrams, the result is proportional to
\bea
\propto & \int_{-\infty}^{\infty} dy^- \int_{-\infty}^{\infty} dy_1^- \int_{-\infty}^{\infty} dy_2^- e^{i x p^+ y^-} (y_1^- - y_2^-)^2
\langle A|  \bar \psi_q(0) \gamma^+ F_{\sigma}^{~+}(y_2^-) F^{+\sigma}(y_1^-) \psi_q(y^-) |A\rangle
\nnu
&\times
\Big[H_C(\{x_i\},k_T)\theta(y_1^- - y^-) \theta(y_2^-) - H_L(\{x_i\},k_T)\theta(y_1^- - y_2^-) \theta(y_2^-)
\nnu
&
- H_R(\{x_i\},k_T)\theta(y_1^- - y^-) \theta(y_2^- - y_1^-)\Big]_{k_T\to 0}.
\label{eq-combine_ss}
\eea
Given that
\bea
H_C(\{x_i\}, k_T=0) = H_L(\{x_i\}, k_T=0) = H_R(\{x_i\}, k_T=0) \equiv H(x,0),
\eea
we have a combination of $\theta$ functions as
\bea
\Big[\theta(y_1^- - y^-) \theta(y_2^-) - \theta(y_1^- - y_2^-) \theta(y_2^-) - \theta(y_1^- - y^-) \theta(y_2^- - y_1^-)\Big],
\eea
which converts Eq.~\eqref{eq-combine_ss} to
\bea
-\int_{-\infty}^{\infty} dy^- e^{i x p^+ y^-} \int_0^{y^-} dy_1^- \int_0^{y_1^-} dy_2^-  (y_1^- - y_2^-)^2
\langle A| \bar \psi_q(0) \gamma^+ F_{\sigma}^{~+}(y_2^-) F^{+\sigma}(y_1^-) \psi_q(y^-) |A\rangle\,
H(x,0).
\eea
In other words, the integration $\int dy_1^- \int dy_2^-$ becomes an ordered integral limited by the value of $y^-$, which is in turn effectively restricted by the rapidly oscillating exponential phase  factor $e^{ixp^+y^-}$, i.e., $y^-\sim 1/xp^+ \to 0$ (if $x$ is not small), and thus also restricts $y_{1,2}^-\to 0$. Physically, this means that all the position integrations in such a term are localized, and therefore, will not have nuclear size enhancement to the double-scattering contribution. These terms (commonly-called ``contact'' terms) can thus be safely neglected when one considers a large nucleus. Therefore, the contributions from soft-soft rescatterings for asymmetric-cut diagrams can be neglected in our calculation for transverse momentum broadening. 

For soft-hard rescattering contributions in left-cut diagrams as shown in Fig.~\ref{fig-left}(b), we follow the same steps in the calculation of soft-soft double scattering in central-cut diagrams, and obtain the following result:
\bea
\frac{d\langle\ell_{hT}^2W^D\rangle^{sh}_L}{dz_h}=&-\frac{2\alpha_s}{N_c}z_h^2(2\pi)^3(1-\epsilon)
\frac{\alpha_s}{2\pi}\int\frac{dx}{x}\int\frac{dz}{z}D_{h/q}(z)\left(\frac{4\pi\mu^2}{Q^2}\right)^{\epsilon}
\frac{1}{\Gamma(1-\epsilon)}\hat{z}^{-\epsilon}(1-\hat{z})^{-\epsilon}\hat{x}^{\epsilon}(1-\hat{x})^{-\epsilon}
\nnu
&\times
\left[x\left.\frac{d}{dx_2}T_{qg}^L(x,x_2,x_B-x)\right|_{x_2\to 0}D_{12}^{sh}+T_{qg}^L(x,0,x_B-x)D_{0}^{sh}\right],
\label{eq-left1}
\eea
where the matrix element $T_{qg}^L$ is given by
\bea
T_{qg}^L(x_1, x_2, x_3)
=&\int \frac{dy^-}{2\pi} e^{ix_1p^+y^-}  \int \frac{dy_1^-dy_2^-}{4\pi} e^{ix_2p^+(y_1^- - y_2^-)}
 e^{ix_3p^+y_2^-} \theta(y_2^-)\theta(y_1^- - y_2^-)
\nnu
&\times  
\langle A|{\bar\psi}_q(0) \gamma^+ F_{\sigma}^+(y_2^-)F^{\sigma +}(y_1^-)\psi_q(y^-)|A\rangle,
\eea
with the $\theta$ functions representing the order of rescatterings. The contribution from soft-hard rescatterings in left-cut diagrams is free of any divergence: the final result denoted as $H_{qg-L}^{sh}\otimes T_{qg}^L$ is given in Eq.~\eqref{eq-HshL} in the Appendix. On the other hand, the contribution from hard-soft rescatterings in the left-cut diagrams in Fig. \ref{fig-left}(c) is zero.

The soft-hard and hard-soft rescatterings in the right-cut diagrams as shown in Fig. \ref{fig-right} are complex conjugates of the ones in the left-cut diagrams, and thus can be obtained directly from the results for diagrams in Fig.~\ref{fig-left}. By replacing the matrix element in Eq. (\ref{eq-HshL})
\bea
T_{qg}^L(x,0,x_B-x)\to T_{qg}^R(x_B,x-x_B,x-x_B),
\eea
with $T_{qg}^R$ given by
\bea
T_{qg}^R(x_1, x_2, x_3)
=&\int \frac{dy^-}{2\pi} e^{ix_1p^+y^-}  \int \frac{dy_1^-dy_2^-}{4\pi} e^{ix_2p^+(y_1^- - y_2^-)}
 e^{ix_3p^+y_2^-} \theta(y_2^--y_1^-)\theta(y_1^- - y^-)
\nnu
&\times  \langle A|{\bar\psi}_q(0) \gamma^+ F_{\sigma}^+(y_2^-)F^{\sigma +}(y_1^-)\psi_q(y^-)|A\rangle,
\eea
we obtain the finite contribution from hard-soft rescatterings at right cut, denoted by $H_{qg-R}^{hs}\otimes T_{qg}^R$, as given in Eq.~(\ref{eq-HhsR}). Similarly, the finite contribution in soft-hard rescatterings at right cut is zero.

Combining all contributions from asymmetric-cut diagrams, the final result is free of any divergence,
\bea
\frac{d\langle\ell_{hT}^2\sigma^D\rangle^{\rm (A)}}{d{\cal PS}}=&-\sigma_h\frac{\alpha_s}{2\pi}\sum_qe_q^2\int\frac{dx}{x}\int\frac{dz}{z}D_{h/q}(z) H_{qg}^{A}\otimes T_{qg}^A,
\eea
where $H_{qg}^{A}\otimes T_{qg}^A$ is given by
\bea
H_{qg}^{A}\otimes T_{qg}^A = H_{qg-L}^{sh}\otimes T_{qg}^L + H_{qg-R}^{hs}\otimes T_{qg}^R.
\label{finite-A}
\eea
with the two terms on the right-hand side given by Eqs.~\eqref{eq-HshL} and \eqref{eq-HhsR}, respectively. Note again that longitudinal contributions for asymmetric-cut diagrams vanish.

\subsection{Gluon-gluon double scattering}

In this subsection, we consider the gluon-gluon double-scattering process in SIDIS, as shown in 
Fig.~\ref{fig-central-gg}, where two initial gluons in the nucleus participate in the process and the first hard gluon plays the same role as the hard quark in the quark-gluon double scattering. Here, for simplicity, we only consider the situation where a quark fragments into the final-state observed hadron. The inclusion of the antiquark fragmentation is made straightforward, by simply replacing the fragmentation function $D_{h/q}(z)\to D_{h/\bar q}(z)$.
\bef
\psfig{file=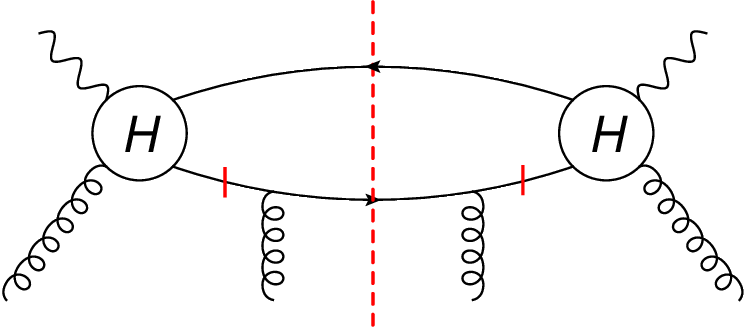, width=2in}
\caption{The central-cut diagram for soft-soft gluon-gluon double scatterings in SIDIS. The short bars indicate the propagators where the soft poles arise. The blob with ``$H$'' inside represents the hard $2\to 2$ processes as shown in Fig. \ref{fig-photon-gluon}.}
\label{fig-central-gg}
\eef
\bef
\psfig{file=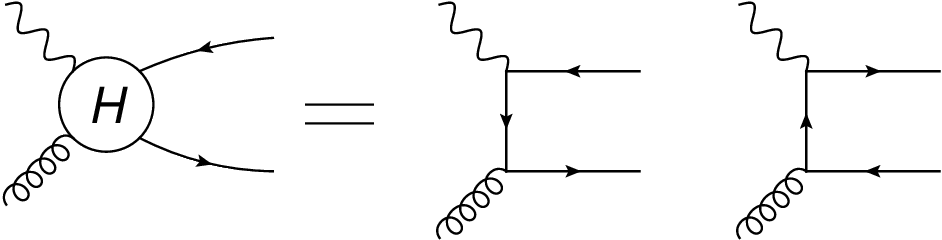, width=3in}
\caption{The representation of hard $2\to 2$ processes for photon-gluon interaction. }
\label{fig-photon-gluon}
\eef

In gluon-gluon double scattering, we only have soft-soft double scattering as illustrated in central-cut diagrams in Fig.~\ref{fig-central-gg}. The kinematics and pole structures are exactly the same as those in soft-soft process of quark-gluon double scattering. The calculation is straightforward and the final result turns out to be
\bea
\frac{d\langle\ell_{hT}^2W^D\rangle^{ss}_{gg}}{dz_h}=&\frac{2\alpha_s}{N_c}z_h^2(2\pi)^3(1-\epsilon)\frac{\alpha_s}{2\pi}\int\frac{dx}{x}\int\frac{dz}{z}D_{h/q}(z)\left(\frac{4\pi\mu^2}{Q^2}\right)^{\epsilon}
\frac{1}{\Gamma(1-\epsilon)}\hat{z}^{-\epsilon}(1-\hat{z})^{-\epsilon}\hat{x}^{\epsilon}(1-\hat{x})^{-\epsilon}
\nnu
&\times\left[x^2\frac{d^2}{dx^2}T_{gg}(x,0,0)D_{2}^{ss}+x\frac{d}{dx}T_{gg}(x,0,0)D_{1}^{ss}+T_{gg}(x,0,0)D_{0}^{ss}\right],
\label{gg-term}
\eea
where the gluon-gluon matrix element $T_{gg}(x, 0, 0)$ is given by \cite{Kang:2011bp}
\bea
T_{gg}(x, 0, 0) =& \frac{1}{x p^+} \int \frac{dy^{-}}{2\pi}\, e^{ix p^{+}y^{-}}
 \int \frac{dy_1^{-}dy_{2}^{-}}{2\pi}  \theta(y_{2}^{-})\,  \theta(y_1^{-}-y^{-}) 
\langle A| F_\alpha^{~+}(0)
F^{\sigma+}(y_2^-)F^+_{~\sigma}(y_1^-)F^{+\alpha}(y^-)|A\rangle\, .
\label{eq-Tgg}
\eea

The $\epsilon$ expansion in Eq.~\eqref{gg-term} gives
\bea
I\times D_{2}^{ss}=&-\frac{1}{\epsilon}T_R\delta(1-\hat z)(1-\hat x)^2(2\hat x^2-2\hat x+1)+\cdots,
\\
I\times D_{1}^{ss}=&\frac{1}{\epsilon}T_R\delta(1-\hat z)(1-\hat x)(1-2\hat x)(6\hat x^2-6\hat x+1)+\cdots,
\\
I\times D_{0}^{ss}=&-\frac{1}{\epsilon}T_R\delta(1-\hat z)(1-\hat x)(1-4\hat x)(6\hat x^2-6\hat x+1)+\cdots,
\eea
which leads to the following divergent piece:
\bea
T_R\int_{x_B}^1\frac{dx}{x}T_{gg}(x,0,0)\left[-\frac{1}{\epsilon}\delta(1-\hat z)(2\hat x^2-2\hat x+1)\right]
=\left(-\frac{1}{\epsilon}\right) \delta(1-\hat z) \int_{x_B}^1\frac{dx}{x}T_{gg}(x,0,0) P_{qg}(\hat x).
\eea 
The remaining finite contribution from gluon-gluon double scattering, denoted as $H_{gg}^C\otimes T_{gg}$, is given in Eq.~(\ref{finite-gg-C}). 

One should in principle also include the asymmetric-cut diagrams in gluon-gluon double scattering. However, these diagrams can be neglected due to the lack of nuclear enhancement as they lead to a contact contribution. Thus the gluon-gluon double-scattering contribution can be written as
\bea
\frac{d\langle\ell_{hT}^2\sigma^D\rangle^{\rm gg}}{d{\cal PS}}=&\sigma_h\frac{\alpha_s}{2\pi}\sum_qe_q^2\int\frac{dx}{x}\int\frac{dz}{z}D_{h/q}(z)\left(\frac{4\pi\mu^2}{Q^2}\right)^{\epsilon}
\frac{1}{\Gamma(1-\epsilon)}
\bigg[ -\frac{1}{\epsilon} \delta(1-\hat z) P_{qg}(\hat x) T_{qg}(x,0,0)
+H_{gg}^{C}\otimes T_{gg}\bigg].
\eea
which comes from the central-cut diagrams only, and $H_{gg}^{C}\otimes T_{gg}$ is expressed in Eq. (\ref{finite-gg-C}).

\subsection{Final result and QCD evolution equation for quark-gluon correlation function}
With all the real and virtual corrections given in the previous subsections, we can combine them and present the final result of transverse momentum broadening in SIDIS at NLO.  Here the real corrections include both central-cut and asymmetric-cut diagrams for the quark-gluon correlation function, and the central-cut diagrams for the gluon-gluon correlation function. We show that all the soft divergences cancel out between real and virtual diagrams. This is an important check for any calculation within the collinear factorization formalism \cite{soft-divergence}. The remaining collinear divergences can be absorbed by the redefinition of either the quark fragmentation function or the quark-gluon correlation function. 

First, let us concentrate on the double-pole $1/\epsilon^2$ terms, which represent soft-collinear divergences. We find that they cancel out between real and virtual contributions [see in particular, the real contribution from the quark-gluon correlation function in Eq.~\eqref{eq-cfinal} and the virtual correction in Eq.~\eqref{eq-virtual}]. Thus we are left with only the $1/\epsilon$ divergences and the finite terms, and they can be written as
\bea
\frac{d\langle \ell_{hT}^2\sigma^D\rangle}{d{\cal PS}}  = &
\sigma_h \frac{\alpha_s}{2\pi}\sum_q e_q^2 \int\frac{dz}{z} D_{h/q}(z)\int\frac{dx}{x} 
\Bigg\{\left(-\frac{1}{\hat\epsilon} + \ln\frac{Q^2}{\mu^2} \right)
\Big[\delta(1-\hat x) P_{qq}(\hat z) T_{qg}(x,0,0) 
+ \delta(1-\hat z)\Big({\mathcal P}_{qg\to qg} \otimes T_{qg} 
\nnu
&
+ P_{qg}(\hat x)T_{gg}(x,0,0) \Big)\Big] 
+ H_{qg}^{C-R} \otimes T_{qg} + H_{qg}^{C-V} \otimes T_{qg}
- H_{qg}^{A} \otimes T_{qg}^A
+ H_{gg}^{C} \otimes T_{gg}
\Bigg\},
\label{Eq-totdiv}
\eea
where the finite corrections in the second line are given by Eqs.~\eqref{finite-C-R}, \eqref{finite-virtual}, \eqref{finite-A}, and \eqref{finite-gg-C}, respectively, $1/\hat\epsilon=1/\epsilon-\gamma_E+\ln(4\pi)$, and $P_{qq}(\hat z)$ and $P_{qg}(\hat x)$ are the usual quark-to-quark and gluon-to-quark splitting kernels as given in Eqs.~\eqref{Pqq} and \eqref{Pqg}, respectively. There is a new term ${\mathcal P}_{qg\to qg} \otimes T_{qg}$ defined as,
\bea
{\mathcal P}_{qg\to qg} \otimes T_{qg}  \equiv  &
P_{qq}(\hat x) T_{qg}(x, 0, 0) 
+ \frac{C_A}{2} \bigg\{ \frac{4}{(1-\hat x)_+} 
T_{qg}(x_B, x-x_B, 0) - \frac{1+\hat x}{(1-\hat x)_+}
\big[T_{qg}(x,0,x_B-x)
\nnu
&
+T_{qg}(x_B,x-x_B,x-x_B)
\big]\bigg\}
+2C_A\delta(1-\hat x)T_{qg}(x,0,0).
\eea
It is thus obvious that the first term that is proportional to $\delta(1-\hat x)$ in Eq.~\eqref{Eq-totdiv} amounts to just the leading-twist collinear QCD correction to the leading-order quark-to-hadron fragmentation function $D_{h/q}(z_h)$:
\bea
D_{h/q}(z_h,\mu_f^2)=D_{h/q}(z_h)-\frac{\alpha_s}{2\pi}\left(\frac{1}{\hat\epsilon}+\ln\frac{\mu^2}{\mu_f^2}\right)\int_{z_h}^1\frac{dz}{z}P_{qq}(\hat z)D_{h/q}(z),
\eea
where we have adopted the $\overline{\rm MS}$ scheme, and $\mu_f$ is the factorization scale for the fragmentation function. The factorization scale $\mu_f$ dependence leads to the same DGLAP evolution equation for the fragmentation function $D_{h/q}(z_h,\mu_f^2)$ as in the single-scattering (leading-twist) case.

Following the same procedure of collinear factorization, one can absorb the second collinear divergence which is proportional $\delta(1-\hat z)$) in Eq.~\eqref{Eq-totdiv} into the redefinition of the corresponding quark-gluon correlation function $T_{qg}(x_B, 0, 0)$,
\bea
T_{qg}(x_B, 0, 0, \mu_f^2)  =  T_{qg}(x_B,0,0) -\frac{\alpha_s}{2\pi}\left(\frac{1}{\hat\epsilon}+\ln\frac{\mu^2}{\mu_f^2}\right)\int_{x_B}^1\frac{dx}{x}
\Big[{\mathcal P}_{qg\to qg} \otimes T_{qg}+P_{qg}(\hat x)T_{gg}(x,0,0)\Big],
\eea 
where we have chosen the same factorization scale $\mu_f$ as that in the fragmentation function. In principle, they do not have to be the same. The above redefinition leads to a new QCD evolution equation for the ``diagonal'' quark-gluon correlation function:
\bea
\mu_f^2 \frac{\partial}{\partial \mu_f^2} T_{qg}(x_B,0,0,\mu_f^2)  =  \frac{\alpha_s}{2\pi} 
\int_{x_B}^1 \frac{dx}{x} \Big[{\mathcal P}_{qg\to qg} \otimes T_{qg} 
+ P_{qg}(\hat x) T_{gg}(x, 0, 0, \mu_f^2)\Big].
\label{eq-evolution}
\eea
This evolution equation, as it stands, is not closed. It is a common feature for higher-twist parton distributions~\cite{Kang:2012ns,VITEV:2014wza,Vogelsang:2009pj}. Under certain approximations for the functional form in $x_{i=1, 2, 3}$ of the two-parton correlation functions, one could obtain a solution to the above evolution equation \cite{Osborne:2002st,Kang:2008ey}.  According to the analysis of induced gluon spectra in Refs.~\cite{Wang:2001ifa,Guo:2000nz,Xing:2011fb,Xing:2012nn}, the interference between soft and hard contributions corresponds to Landau-Pomeranchuk-Migdal (LPM) \cite{Migdal:1956tc} interference which suppresses gluon radiation with a large  formation time $\tau_f\gg R_A$. In the high-twist formalism, the formation time of the medium-induced gluon is defined as $\tau_f=1/[x(1-\hat x)p^+]$. Thus the LPM region can be reached if we impose $\hat x\to 1$ in Eq.~\eqref{eq-evolution}. In this particular kinematic region, the interference between soft and hard rescatterings gives rise to a destructive effect to the final contribution.

Notice that in Eq.~\eqref{eq-evolution}, the quark-gluon correlation function $T_{qg}$ is coupled with the gluon-gluon correlation function. In order to solve the evolution equation, in principle, one needs an evolution equation for the gluon-gluon correlation function $T_{gg}$.  However, since we deal with the SIDIS process, in which only $T_{qg}$ enters at the LO as in Eq.~\eqref{LO-weight}, our NLO calculation cannot give the evolution equation for $T_{gg}$. In this case, we have to go beyond NLO to NNLO, or we could study transverse momentum broadening for a process where $T_{gg}$ enters at the LO, e.g., a scalar particle production in the gluon-gluon fusion channel in proton-nucleus ($p+A$) collisions. This way we will be able to derive the complete set of evolution equations which couple both $T_{qg}$ and $T_{gg}$ correlation functions. 

Under the approximation of a large and loosely bound nucleus where one can neglect the momentum and spatial correlations of two nucleons \cite{Osborne:2002st}, we can express the quark-gluon correlation function $T_{qg}(x_B, 0, 0, \mu_f^2)$ in a factorized form \cite{CasalderreySolana:2007sw},
\bea
T_{qg}(x_B,0,0,\mu_f^2) \approx \frac{N_c}{4\pi^2\alpha_{\rm s}} f_{q/A}(x_B, \mu_f^2) \int dy^- \hat {q}(\mu_f^2,y^-),
\eea
where $f_{q/A}(x_B, \mu_f^2)$ is the standard quark distribution function inside a nucleus and $\hat {q}(\mu_f^2,y^-)$ is the jet transport parameter that describes the averaged transverse momentum transfer squared per unit distance (or mean free path) in the medium. Thus from Eq.~\eqref{eq-evolution} one could in principle determine the factorization scale $\mu_f^2$ dependence of $\hat {q}(\mu_f^2,y^-)$. Such QCD evolution of $\hat {q}(\mu_f^2,y^-)$ will have important consequences on quantitative studies of jet quenching at NLO. A preliminary study for transverse momentum broadening based on such a formalism with $\hat {q}(\mu_f^2,y^-)$ evolution incorporated shows good agreement with experimental data from both $e+A$ and $p+A$ collisions \cite{fitting}.

After $\overline{\rm MS}$ subtraction of the collinear divergences into the fragmentation function $D_{h/q}(z,\mu_f^2)$ and the twist-4 quark-gluon correlation function $T_{qg}(x, 0, 0, \mu_f^2)$, we can express the $\ell_{hT}^2$-weighted differential cross section up to NLO at twist 4 as,
\bea
\frac{d\langle \ell_{hT}^2\sigma^D\rangle}{d{\cal PS}} =&
\sigma_h\sum_qe_q^2\int_{x_B}^1 \frac{dx}{x}T_{qg}(x,0,0,\mu_f^2)\int_{z_h}^1\frac{dz}{z}D_{h/q}(z,\mu_f^2)\delta(1-{\hat x})\delta(1-{\hat z})
\nnu
&
+\sigma_h\frac{\alpha_s}{2\pi} \sum_q e_q^2 \int_{z_h}^1 \frac{dz}{z}D_{h/q}(z, \mu_f^2) 
\int_{x_B}^1\frac{dx}{x} \bigg\{\ln\left(\frac{Q^2}{\mu_f^2}\right)  
\Big[\delta(1-\hat x)P_{qq}(\hat z) T_{qg}(x, 0, 0, \mu_f^2)
\nnu
&
+ \delta(1-\hat z) \big({\mathcal P}_{qg\to qg}\otimes T_{qg} 
+ P_{qg}(\hat x) T_{gg}(x, 0, 0, \mu_f^2)\big) \Big]
\nnu
&
+ H_{qg}^{C-R} \otimes T_{qg} + H_{qg}^{C-V} \otimes T_{qg}
- H_{qg}^{A} \otimes T_{qg}^A
+ H_{gg}^{C} \otimes T_{gg}\bigg\},
\label{eq-double}
\eea
which includes finite NLO corrections to the hard-part coefficient function. Just like the NLO correction to the differential cross section at leading twist in Eq.~\eqref{eq-single}, the finite hard part coefficient at NLO also depends on the factorization scale which will reduce the overall factorization scale dependence of the cross section when combined with the scale dependence of the fragmentation functions and the twist-4 quark-gluon correlation function as determined by the evolution equations. Substituting the $\ell_{hT}^2$-weighted cross section in Eq.~\eqref{eq-double} and leading-twist differential cross section in Eq.~\eqref{eq-single} into Eq.~\eqref{eq-broadening definition}, we are able to compute the transverse momentum broadening in SIDIS at NLO, which is the main result of this paper.  

Our results in this paper verify for the first time the factorization of the $\ell_{hT}^2$-weighted differential cross section at twist 4 in NLO. The collinear divergences associated with the quark fragmentation function and twist-4 quark-gluon correlation function are factorized, and one is left only with finite hard coefficient functions, which also depend on the factorization scale.  One should also consider contributions from double quark scattering \cite{Schafer:2007xh} and hadron production from gluon fragmentation for more complete NLO calculations. These will be left for future publications. 

It is also worth mentioning that the techniques we have developed here in principle can also be applied to study the situation where one uses a different weighting factor instead of $\ell_{hT}^2$. One of such possibilities will be a Bessel weighting as advocated in Ref. \cite{Boer:2011xd}. For the advantages of such a Bessel weighting, see, e.g. Refs.~\cite{Boer:2015kxa,Boer:2014bya}. We plan to study such a possibility in the future.

\section{Summary}
We have calculated the NLO pQCD corrections to the nuclear transverse momentum broadening in semi-inclusive hadron production in deep inelastic $e+A$ collisions. Specifically, we have demonstrated in detail how to evaluate at NLO the transverse-momentum-weighted differential cross section at twist 4. By including contributions from quark-gluon and gluon-gluon double scatterings, as well as interferences between single and triple scatterings, we have shown explicitly that soft divergences cancel out between real and virtual corrections, and the remaining collinear divergences can be absorbed into the redefinition (renormalization) of the final-state fragmentation function and initial-state twist-4 quark-gluon correlation function, which enabled us to identity a DGLAP-type evolution equations for the twist-4 quark-gluon correlation function. After the subtraction of collinear divergences, the transverse-momentum-weighted cross section can be factorized as a convolution of twist-4 nuclear parton correlation functions, the usual twist-2 fragmentation function and hard parts which are finite and free of any divergence. With the NLO results for inclusive cross section and transverse-momentum-weighted differential cross section in hand, our result can be further applied to phenomenological studies of transverse momentum broadening in HERMES and experiments at the Jefferson Lab experiments and future Electron-Ion Collider facilities. Such detailed phenomenological studies will be carried out in a forthcoming paper \cite{fitting}.

We want to emphasize that it is important to perform similar studies for some other processes. For example, through the NLO calculations of transverse momentum broadening in Drell-Yan lepton pair production in $p+A$ collisions, we can verify the collinear factorization at twist 4, and demonstrate the universality of the twist-4 quark-gluon correlation function. This will be published in a separate paper. On the other hand, an extension to a scalar particle production through the gluon-gluon fusion channel in $p+A$ collisions will enable us to study the evolution equation for the twist-4 gluon-gluon correlation function, from which we can derive a complete set of evolution equations for twist-4 parton correlation functions.

\section*{Acknowledgments}
This work is supported by the U.S. Department of Energy, Office of Science, Office of High Energy and Nuclear Physics, Division of Nuclear Physics, under Contract No.~DE-AC52-06NA25396 and No. DE-AC02-05CH11231, and within the framework of the JET Collaboration,  the National Science Foundation of China under Grants No.~11221504 and No. 10825523, China Ministry of Science and Technology under Grant No. 2014DFG02050, and the Major State Basic Research Development Program in China (No. 2014CB845404).

\appendix
\section{Complete list of finite terms}
In this appendix, we list the finite terms in the leading-twist differential cross section and twist-4 weighted differential cross section at NLO. The finite terms $H^{NLO}_{T2-qq}$, $H^{NLO}_{T2-qg}$, and $H^{NLO}_{T2-gq}$ for the leading-twist differential cross section at NLO in Eq.~\eqref{eq-single} can be written as
\bea
H^{NLO}_{T2-qq} =& C_F\bigg\{-8\delta(1-\hx)\delta(1-\hz)
+\frac{1+(1-\hx-\hz)^2}{(1-\hx)_+(1-\hz)_+}
+\delta(1-\hz)\left[(1+\hx^2) \left(\frac{\ln(1-\hx)}{1-\hx}\right)_+ 
-\frac{1+\hx^2}{1-\hx}\ln\hx+(1-\hx)
\right]
\nnu
&
+\delta(1-\hx) \left[(1+\hz^2) \left(\frac{\ln(1-\hz)}{1-\hz}\right)_+ 
+\frac{1+\hz^2}{1-\hz}\ln\hz+(1-\hz)
\right]
+\frac{1+4(1-y)+(1-y)^2}{1+(1-y)^2}2\hat x \hat z\bigg\},
\label{eq-finiteT2-qq}
\\
H^{NLO}_{T2-qg}=&
\ln \Big[\hat z (1-\hat z)\Big] P_{gq}(\hat z)\delta(1-\hat x)
+C_F\left[\frac{1+(\hat x-\hat z)^2}{\hat z(1-\hat x)_+}
+\hat z\delta(1-\hat x)
+\frac{1+4(1-y)+(1-y)^2}{1+(1-y)^2}2\hat x(1-\hat z)\right],
\label{eq-finiteT2-qg}
\\
H^{NLO}_{T2-gq}=&
\ln\frac{1-\hat x}{\hat x} P_{qg}(\hx)\delta(1-\hz)
+T_R\bigg[\frac{2\hat x^2-2\hat x+2\hat z^2-2\hat z+1}{\hat z(1-\hat z)_+}
+2\hat x(1-\hat x)\delta(1-\hat z)+\frac{1+4(1-y)+(1-y)^2}{1+(1-y)^2}
\nnu
&\times 4\hat x (1-\hat x)\bigg].
\label{eq-finiteT2-gq}
\eea

For the twist-4 weighted differential cross section, besides the finite term for virtual diagrams as given in Eq.~\eqref{finite-virtual}, there are nine finite terms. For the central-cut diagrams, we have four finite terms:  $H_{qg-C}^{ss}\otimes T_{qg}$ associated with soft-soft double scattering, $H_{qg-C}^{hh}\otimes T_{qg}$ associated with hard-hard double scattering, $H_{qg-C}^{sh}\otimes T_{qg}$ associated with soft-hard double scattering, and $H_{qg-C}^{hs}\otimes T_{qg}$ associated with hard-soft double scatterings. For the asymmetric-cut diagrams, we also have four finite terms: 
$H_{qg-L}^{sh}\otimes T_{qg}^L$ (or $H_{qg-L}^{sh}\otimes T_{qg}^R$) associated with soft-hard scattering in left-cut (right-cut) diagrams, $H_{qg-L}^{hs}\otimes T_{qg}^L$ (or $H_{qg-L}^{hs}\otimes T_{qg}^R$) associated with hard-soft scattering in left-cut (right-cut)diagrams. At the same time, we also have the finite term for gluon-gluon double scattering $H_{gg}^C\otimes T_{gg}$. They are given by the following expressions:
\bea
H_{qg-C}^{ss}\otimes T_{qg}=&x^2\frac{d^2}{dx^2}T_{qg}(x,0,0)C_F\Bigg\{
\frac{(1-\hat x)(\hat x^2+2\hat x\hat z-2\hat x+\hat z^2-2\hat z+2)}{\hat z^2(1-\hat z)_+}
\nnu
&
-\delta(1-\hat z)(1-\hat x)\left[2\hat x+\ln\frac{\hat x}{1-\hat x}(1+\hat x^2)\right] +\frac{1+4(1-y)+(1-y)^2}{1+(1-y)^2}\frac{2\hat x(1-\hat x)^2}{\hat z}\Bigg\}
\nnu
&
-x\frac{d}{dx}T_{qg}(x,0,0)C_F\Bigg\{
\frac{-4\hat x^3+\hat x^2(9-4\hat z)-6\hat x(1-\hat z)+(\hat z-2)\hat z+2}{\hat z^2(1-\hat z)_+}
\nnu
&
+\delta(1-\hat z)\bigg[(3\hat x^2-6\hat x-1) + \ln\frac{\hat x}{1-\hat x}(4\hat x^3-5\hat x^2-1)\bigg]
-\frac{1+4(1-y)+(1-y)^2}{1+(1-y)^2}\frac{2\hat x(1-\hat x)(3-4\hat x)}{\hat z}
\Bigg\}
\nnu
&
+T_{qg}(x,0,0)C_F\Bigg\{
\frac{2\hat x\hat z(2\hat x^2-5\hat x+4)+\hat x^2(6\hat x^2-18\hat x+19)-8\hat x+(1-\hat z)^2+1}{\hat z^2(1-\hat x)_+(1-\hat z)_+}
+4\delta(1-\hat x)\delta(1-\hat z)
\nnu
&
+\delta(1-\hat z)\left[\left(\frac{\ln(1-\hat x)}{1-\hat x}\right)_+ - \frac{\ln\hat x}{1-\hat x}\right]\left[1+\hat x^2(6\hat x^2-14\hat x+9)\right]
\nnu
&
-\delta(1-\hat z) \frac{2\hat x^3-7\hat x^2+8\hat x+1}{(1-\hat x)_+}
+\delta(1-\hat x)\left[\left(\frac{\ln(1-\hat z)}{1-\hat z}\right)_+ + \frac{\ln\hat z}{1-\hat z}\right]\frac{1+\hat z^2}{\hat z^2}
\nnu
&
-\delta(1-\hat x)\frac{(1+\hat z)^2}{\hat z^2(1-\hat z)_+}
+\frac{1+4(1-y)+(1-y)^2}{1+(1-y)^2}\frac{4\hat x(1-\hat x)(2-3\hat x)}{\hat z}
\Bigg\},
\label{eq-finite-ss}
\\
H_{qg-C}^{hh}\otimes T_{qg}=&T_{qg}(x_B,x-x_B,0)C_A\left\{
\delta(1-\hat x)\left[\left(\frac{\ln(1-\hat z)}{1-\hat z}\right)_+ + \frac{\ln\hat z}{1-\hat z}\right]\frac{(1+\hat z^2)\left[C_F/C_A(1-\hat z)^2+\hat z\right]}{\hat z^2}\right.
\nnu
&
+\delta(1-\hat x)\frac{(1-\hat z)\left[C_F/C_A(1-\hat z)^2+\hat z\right]}{\hat z^2}
+\frac{(1+\hat z^2)\left[C_F/C_A(1-\hat z)^2+\hat z\right]}{\hat z^2(1-\hat x)_+(1-\hat z)_+}
\nnu
&
\left.+2\delta(1-\hat z)\left[\left(\frac{\ln(1-\hat x)}{1-\hat x}\right)_+ - \frac{\ln\hat x}{1-\hat x}\right]
\right\},
\label{eq-finite-hh}
\\
H_{qg-C}^{sh}\otimes T_{qg}=&
x\frac{d}{dx}T_{qg}(x,0,x_B-x)\frac{C_A}{2}\Bigg\{
\frac{(1+\hat x\hat z^2)[\hat z+2C_F/C_A(1-\hat z)]}{\hat z^2(1-\hat z)_+}
-\delta(1-\hat z)\left(1+\hat x\right)\left(1+\ln\frac{\hat x}{1-\hat x}\right)\Bigg\}
\nnu
&
-x\left.\frac{d}{dx_2}T_{qg}(x,x_2,x_B-x)\right|_{x_2\to 0}
\frac{C_A}{2}\left(\frac{1}{\hat z^2}+\hat x\right)\left[\hat z+2C_F/C_A(1-\hat z)\right]
\nnu
&
+T_{qg}(x,0,x_B-x)\frac{C_A}{2}\Bigg\{
\frac{(\hat x^2\hat z^2-2\hat x\hat z^2-1)[\hat z+2C_F/C_A(1-\hat z)]}{\hat z^2(1-\hat x)_+(1-\hat z)_+}
-2\delta(1-\hat x)\delta(1-\hat z)
\nnu
&
+\delta(1-\hat z)\left[\left(\frac{\ln(1-\hat x)}{1-\hat x}\right)_+ - \frac{\ln\hat x}{1-\hat x}\right]\left(\hat x^2-2\hat x-1\right)
-\delta(1-\hat z)\frac{\hat x^2-3\hat x-2}{(1-\hat x)_+}
\nnu
&
-\delta(1-\hat x)\left[\left(\frac{\ln(1-\hat z)}{1-\hat z}\right)_+ + \frac{\ln\hat z}{1-\hat z}\right]\frac{1+\hat z^2}{\hat z^2}[\hat z+2C_F/C_A(1-\hat z)]
\nnu
&
+\delta(1-\hat x)\frac{(2\hat z^3-\hat z^2+4\hat z-1)\left[\hat z+2C_F/C_A(1-\hat z)\right]}{\hat z^2(1-\hat z)_+}
\Bigg\},
\label{eq-finite-sh}
\\
H_{qg-C}^{hs}\otimes T_{qg}=&
x\frac{d}{dx}T_{qg}(x_B,x-x_B,x-x_B)\frac{C_A}{2}\Bigg\{
\frac{(1+\hat x\hat z^2)[\hat z+2C_F/C_A(1-\hat z)]}{\hat z^2(1-\hat z)_+}
-\delta(1-\hat z)\left(1+\hat x\right)\left(1+\ln\frac{\hat x}{1-\hat x}\right)\Bigg\}
\nnu
&
-x\left.\frac{d}{dx_2}T_{qg}(x_B,x_2,x-x_B)\right|_{x_2\to x-x_B}
\frac{C_A}{2}\left(\frac{1}{\hat z^2}+\hat x\right)\left[\hat z+2C_F/C_A(1-\hat z)\right]
\nnu
&
+T_{qg}(x_B,x-x_B,x-x_B)\frac{C_A}{2}\Bigg\{
\frac{(\hat x^2\hat z^2-2\hat x\hat z^2-1)[\hat z+2C_F/C_A(1-\hat z)]}{\hat z^2(1-\hat x)_+(1-\hat z)_+}
-2\delta(1-\hat x)\delta(1-\hat z)
\nnu
&
+\delta(1-\hat z)\left[\left(\frac{\ln(1-\hat x)}{1-\hat x}\right)_+ - \frac{\ln\hat x}{1-\hat x}\right]\left(\hat x^2-2\hat x-1\right)
-\delta(1-\hat z)\frac{\hat x^2-3\hat x-2}{(1-\hat x)_+}
\nnu
&
-\delta(1-\hat x)\left[\left(\frac{\ln(1-\hat z)}{1-\hat z}\right)_+ + \frac{\ln\hat z}{1-\hat z}\right]\frac{1+\hat z^2}{\hat z^2}[\hat z+2C_F/C_A(1-\hat z)]
\nnu
&
+\delta(1-\hat x)\frac{(2\hat z^3-\hat z^2+4\hat z-1)\left[\hat z+2C_F/C_A(1-\hat z)\right]}{\hat z^2(1-\hat z)_+}
\Bigg\},
\label{eq-finite-hs}
\\
H_{qg-L}^{sh}\otimes T_{qg}^L=&-x\left.\frac{d}{dx_2}T_{qg}^L(x,x_2,x_B-x)\right|_{x_2\to 0}\frac{C_A}{2}
\left(\frac{1}{\hat z^2}+\hat x\right)\left[\hat z+2C_F/C_A(1-\hat z)\right]
\nnu
&
-T_{qg}^L(x,0,x_B-x)\delta(1-\hat x)C_A\left(\frac{1}{\hat z^2}+1\right)\left[\hat z+2C_F/C_A(1-\hat z)\right]
\label{eq-HshL}
\\
H_{qg-R}^{hs}\otimes T_{qg}^R=&-x\left.\frac{d}{dx_2}T_{qg}^R(x_B,x_2,x-x_B)\right|_{x_2\to x-x_B}\frac{C_A}{2}
\left(\frac{1}{\hat z^2}+\hat x\right)\left[\hat z+2C_F/C_A(1-\hat z)\right]
\nnu
&
-T_{qg}^R(x_B,x-x_B,x-x_B)\delta(1-\hat x)C_A\left(\frac{1}{\hat z^2}+1\right)\left[\hat z+2C_F/C_A(1-\hat z)\right]\label{eq-HhsR}
\\
H_{gg}^C\otimes T_{gg}=&x^2\frac{d^2}{dx^2}T_{gg}(x,0,0)T_R\Bigg\{
\frac{(1-\hat x)^2(2\hat x^2-2\hat x+2\hat z^2-2\hat z+1)}{\hat z(1-\hat z)_+}
-\delta(1-\hat z)\ln\frac{\hat x}{1-\hat x}(1-\hat x)^2(2\hat x^2-2\hat x+1)
\nnu
&
-\delta(1-\hat z)(2\hat x^2-3\hat x+1)^2
+\frac{1+4(1-y)+(1-y)^2}{1+(1-y)^2}4\hat x(1-\hat x)^3
\Bigg\}
\nnu
&
-x\frac{d}{dx}T_{gg}(x,0,0)T_R\bigg[
\frac{(1-\hat x)(1-2\hat x)(6\hat x^2-6\hat x+2\hat z^2-2\hat z+1)}{\hat z^3(1-\hat z)_+}
-\delta(1-\hat z)2(1-\hat x)^2(12\hat x^2-7\hat x+1)
\nnu
&
-\delta(1-\hat z)\ln\frac{\hat x}{1-\hat x}(1-\hat x)(1-2\hat x)(6\hat x^2-6\hat x+1)
+\frac{1+4(1-y)+(1-y)^2}{1+(1-y)^2} 12\hat x(2\hat x-1)(1-\hat x)^2\bigg]
\nnu
&
-T_{gg}(x,0,0)T_R\Bigg\{
\frac{(1-\hat x)\left[24\hat x^3-30\hat x^2+2\hat x(2\hat z^2-2\hat z+5)-2\hat z^2+2\hat z-1\right]}{\hat z(1-\hat z)_+}
\nnu
&
+\delta(1-\hat z) \ln\frac{\hat x}{1-\hat x} (1-\hat x) (1-4\hat x)(6\hat x^2-6\hat x+1)
-\delta(1-\hat z)2(1-\hat x)(24\hat x^3-33\hat x^2+11\hat x-1)
\nnu
&
+\frac{1+4(1-y)+(1-y)^2}{1+(1-y)^2}4\hat x(1-\hat x)(12\hat x^2-15\hat x+4)
\Bigg\}.
\label{finite-gg-C}
\eea


\end{document}